\documentclass[aps,twocolumn,superscriptaddress,floatfix]{revtex4-1}
\usepackage[latin9]{inputenc}
\usepackage{amsmath}
\usepackage{amssymb}
\usepackage{graphicx}
\usepackage{esint}
\usepackage{epstopdf}
\usepackage{braket}
\usepackage{color}
\usepackage{hyperref}

\makeatletter
\@ifundefined{textcolor}{}
{
 \definecolor{BLACK}{gray}{0}
 \definecolor{WHITE}{gray}{1}
 \definecolor{RED}{rgb}{1,0,0}
 \definecolor{GREEN}{rgb}{0,1,0}
 \definecolor{BLUE}{rgb}{0,0,1}
 \definecolor{CYAN}{cmyk}{1,0,0,0}
 \definecolor{MAGENTA}{cmyk}{0,1,0,0}
 \definecolor{YELLOW}{cmyk}{0,0,1,0}
}

\makeatother

\begin{document}

\title{Long time universality of black-hole lasers}







\author{J. R. M. de Nova}
\email{jrmnova@fis.ucm.es}
\affiliation{Departamento de F\'isica de Materiales, Universidad Complutense de
Madrid, E-28040 Madrid, Spain}

\author{P. F. Palacios}
\affiliation{Departamento de F\'isica de Materiales, Universidad Complutense de
Madrid, E-28040 Madrid, Spain}

\author{I. Carusotto}
\affiliation{INO-CNR BEC Center and Dipartimento di Fisica, Universit\`{a} di Trento, I-38123 Povo, Italy}

\author{F. Sols}
\affiliation{Departamento de F\'isica de Materiales, Universidad Complutense de
Madrid, E-28040 Madrid, Spain}

\begin{abstract}
For flowing quantum gases, it has been found that at long times an initial black-hole laser (BHL) configuration exhibits only two possible states: the ground state or a periodic self-oscillating state of continuous emission of solitons. So far, all the works on this subject are based on a highly idealized model, quite difficult to implement experimentally. Here we study the instability spectrum and the time evolution of a recently proposed realistic model of a BHL, thus providing a useful theoretical tool for the clear identification of black-hole lasing in future experiments. We further confirm the existence of a well-defined phase diagram at long times, which bespeaks universality in the long-time behavior of a BHL. Additionally, we develop a complementary model in which the same potential profile is applied to a subsonic homogeneous flowing condensate that, despite not forming a BHL, evolves towards the same phase diagram as the associated BHL model. This result reveals an even stronger form of robustness in the long-time behavior with respect to the transient, which goes beyond what has been described in the previous literature.
\end{abstract}
\date{\today}

\maketitle

\section{Introduction}\label{sec:Intro}

The spontaneous emission of radiation by the event horizon of a black hole (BH) \cite{Hawking1974}, known as Hawking radiation, is one of the most celebrated predictions of modern theoretical physics, combining thermodynamics, quantum mechanics, and general relativity. However, its detection in an astrophysical context is unlikely in the foreseeable future due to its extremely low characteristic temperature, much lower than that of the cosmic microwave background. An alternative approach to study such gravitational phenomena is provided by a flowing fluid \cite{Unruh1981}, in which a subsonic-supersonic interface becomes the acoustic analog of an event horizon. Based on this observation, a number of analog setups emerged in systems as different as water waves \cite{WeinfurtnerPRL2011,Euve2016}, nonlinear optical fibers \cite{Belgiorno2010,Drori2019} or quantum fluids of light \cite{RMPCarusotto2013,Nguyen2015}.

Most importantly, Bose-Einstein condensates were revealed as the strongest candidates for the observation of the spontaneous Hawking effect due to their low temperature, well-controlled behavior, and genuine quantumness \cite{Garay2000}. This proposal was studied and expanded in the theoretical literature \cite{Balbinot2008,Carusotto2008,Recati2009,Macher2009,Zapata2011,Larre2012,deNova2014,Finazzi2013,deNova2014a,Busch2014,deNova2015,Michel2016a}, and implemented in parallel in an experimental setup \cite{Lahav2010} which eventually led to the first experimental measurements of spontaneous Hawking radiation and in particular of its entanglement \cite{Steinhauer2016}, its temperature \cite{deNova2019} and its stationarity \cite{Kolobov2021}.

Another interesting analog feature that can be observed in a Bose-Einstein condensate due to its superluminal dispersion relation is the so-called black-hole laser (BHL) effect \cite{Corley1999}, in which a configuration displaying a pair of horizons can give rise to the self-amplification of Hawking radiation due to successive reflections between them, like in a laser cavity, translated into the appearance of dynamical instabilities in the spectrum of excitations \cite{Leonhardt2003,Barcelo2006,Jain2007,Coutant2010,Finazzi2010,Bermudez2018,Burkle2018}. A complete experimental characterization of the black-hole laser effect is still one of the major present challenges in the gravitational analog field, since its first reported observation \cite{Steinhauer2014} has sparked intense discussions within the
community \cite{Tettamanti2016,Steinhauer2017,Wang2016,Wang2017,Llorente2019}. There are also attempts to observe the black-hole laser effect in other analog systems with superluminal dispersion relation \cite{Faccio_2012,Peloquin2016,RinconEstrada2021}.

One of the most intriguing aspects of a black-hole laser is its long-time behavior \cite{Michel2013,Michel2015,deNova2016}, once the initial instability has grown up to saturation and the system enters the full nonlinear regime. This regime provides a unique setup to test backreaction \cite{Balbinot2005a}, in contrast to the case of the usual spontaneous Hawking effect, whose backreaction is too small to be observed in an actual experiment. In particular, it has been seen that the long-time behavior of the system displays a well-defined phase diagram, characterized by solely the parameters of the setup and independent of the details of the transient \cite{deNova2016}. Specifically, for sufficiently long times, the system only exhibits two possible states: either it reaches the true ground state (GS) by evaporating away the horizons or it enters a regime of continuous emission of solitons (CES). Due to its periodic and radiating nature, the CES regime has been argued to be the closest counterpart of an actual optical laser, and the \textit{bona fide} black-hole laser \cite{deNova2016}.

Nevertheless, all the works studying the nonlinear behavior of black-hole lasers \cite{Michel2013,Michel2015,deNova2016} have so far considered the so-called flat-profile model \cite{Carusotto2008,Recati2009}, where the coupling constant and external potential are space dependent and perfectly matched so that the condensate density and flow velocity are homogeneous (``flat''), and only the speed of sound varies. Albeit very useful for analytical calculations due to its simplicity, the flat-profile configuration is quite difficult to implement experimentally. A more realistic approach was provided in Ref. \cite{deNova2017a}, where it was shown that from every single black-hole solution a black-hole laser solution can be constructed. That result was used to propose an adequate theoretical description of the black-hole laser configuration in the actual experiment \cite{Steinhauer2014}, created with the help of an attractive potential well. The advantage of this theoretical model is that it allows to isolate the genuine contribution of black-hole lasing in a setup similar to that used for the real experiment, where this contribution has so far been overshadowed by the Bogoliubov-Cherenkov radiation background \cite{Kolobov2021}.

In this work we present a comprehensive study of the long-time behavior of such realistic model of a black-hole laser, computing also the linear spectrum that describes its unstable behavior at short times. We confirm and extend previous results in the literature for both the linear spectrum and the long-time regime \cite{Michel2013,Michel2015,deNova2016}. In particular, at long times we also find a well-defined phase diagram between GS and CES, which hints at a possible form of long-time universality in the behavior of black-hole lasers.

Furthermore, we develop a novel approach to the study of the long-time behavior of a BHL configuration that yields a rather complete understanding of the problem. Specifically, we analyze the time evolution in a complementary scenario where the same potential profile as before is suddenly applied now to a subsonic homogeneous flowing condensate which, not being in a stationary state of the new potential, evolves deterministically. We find that its long-time phase diagram is identical to that for the BHL solutions in the same setup. This reveals a stronger form of robustness with respect to the transient than previously reported. Namely, it shows that GS and CES are intrinsic states of a flowing condensate that are eventually reached independently of the early-time dynamics.


This paper is arranged as follows. In Sec. \ref{sec:WF} we present the configurations analyzed throughout this work. Section \ref{sec:Preview} is devoted to a preliminary theoretical analysis of the BHL solutions considered in this work, analyzing both their linear and nonlinear spectra. In Sec. \ref{sec:TimeEvolution} we study numerically the time evolution of the initial unstable BHL solutions, verifying what has been anticipated in the previous section. Section \ref{sec:PhaseDiagram} deals with the time evolution of an alternative configuration in which the initial condition is a subsonic homogeneous flowing condensate, and includes the computation of its long-time phase diagram. In Sec. \ref{sec:Robustness} the results of the previous sections are connected and discussed. Conclusions and outlook are presented in Sec. \ref{sec:conclu}.

\section{Analog configurations in flowing condensates}\label{sec:WF}


\begin{figure*}[!htb]
\begin{tabular}{@{}cc@{}}
    \includegraphics[width=\columnwidth]{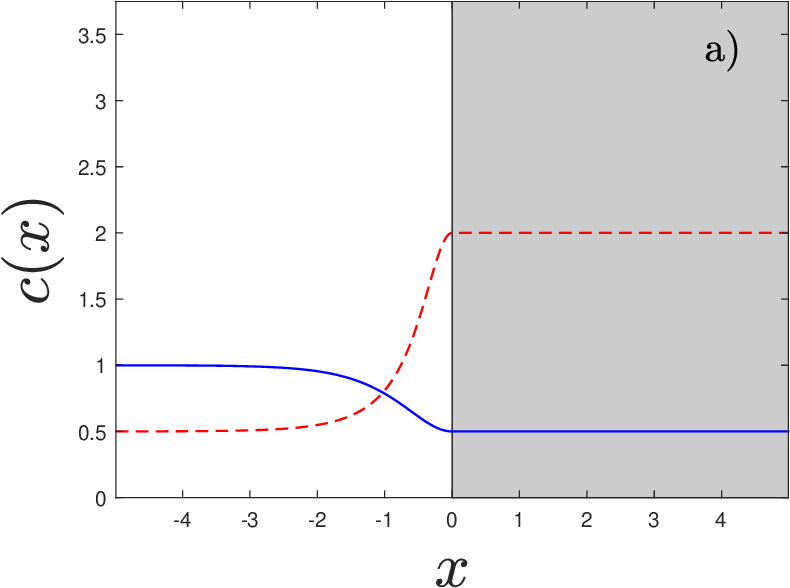}   & \includegraphics[width=\columnwidth]{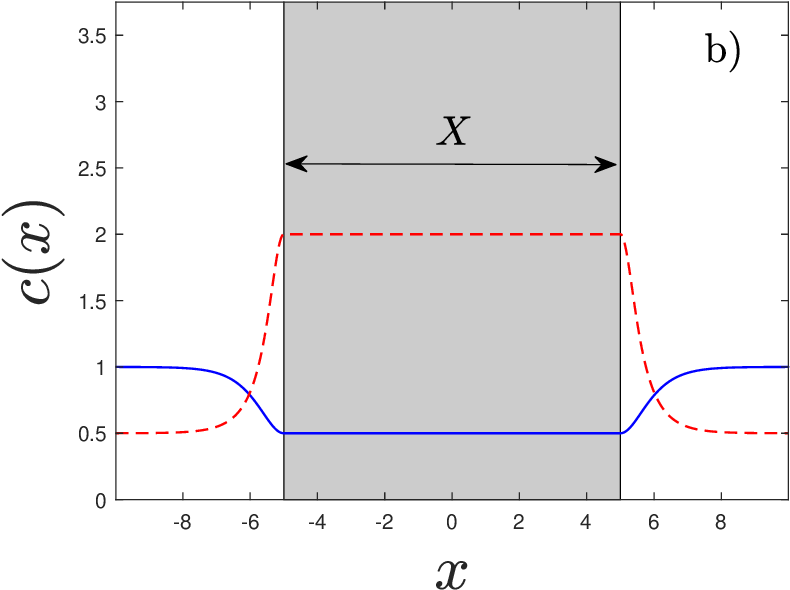} \\[0.5cm]
    \includegraphics[width=\columnwidth]{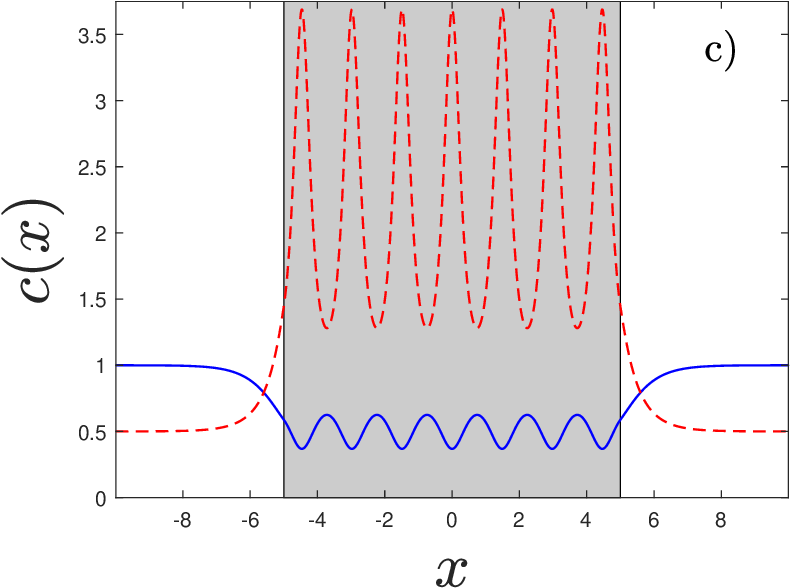} & \includegraphics[width=\columnwidth]{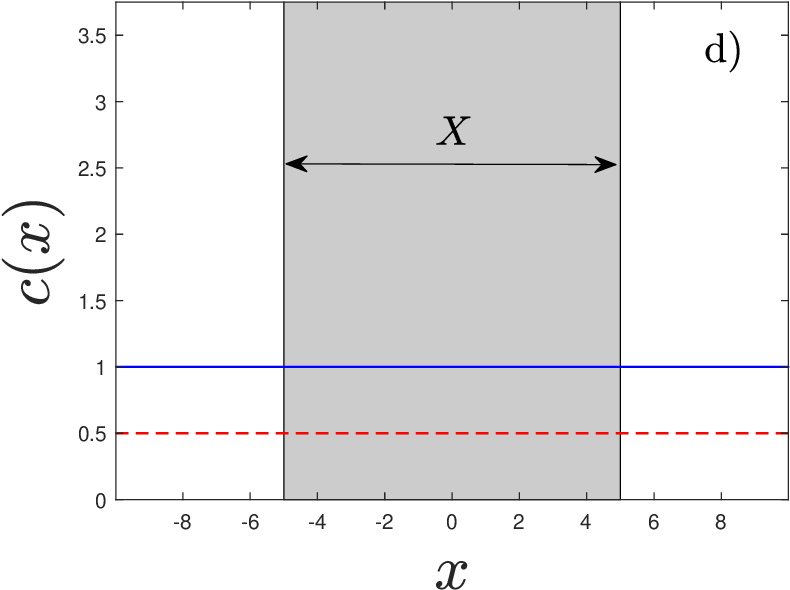}
\end{tabular}
\caption{Spatial profile of sound (solid blue) and flow (dashed red) velocities for different flowing configurations, where the shaded area represents the region in which an attractive constant potential $V(x)=-V_0$ is present. (a) BH waterfall configuration with $v=0.5$, corresponding to a step amplitude $V_0=V_0(v)=1.125$. (b) BHL configuration associated to (a) with $X=10$. (c) Generic symmetric BHL solution for $v=0.5$, $X=10$ and $V_0=1.5$. (d) IHFC with the same background parameters as (b).}
\label{fig:BHWFConfigurations}
\end{figure*}

We devote this section to introduce the main theoretical models used along the work; the interested reader is referred to the literature for the complete technical details \cite{Michel2015,deNova2016,deNova2017a}. We consider a one-dimensional Bose-Einstein condensate close to $T=0$, described by the Gross-Pitaevskii (GP) wave function $\Psi(x,t)$, whose associated sound and flow velocities are
\begin{equation}
    c(x,t)=\sqrt{\frac{gn(x,t)}{m}},~v(x,t)=\frac{\hbar\partial_x\alpha(x,t)}{m}
\end{equation}
where $g$ is the coupling constant, $m$ the mass of the atoms, and we are using a density-phase decomposition of the wave function $\Psi(x,t)=\sqrt{n(x,t)}e^{i\alpha(x,t)}$. Following the usual convention, hereafter we set units such that $\hbar=m=c_u=1$ and rescale the GP wave function as $\Psi(x,t)\rightarrow \sqrt{n_u}\Psi(x,t)$ so it becomes dimensionless, with $n_u,c_u$ the asymptotic subsonic density and speed of sound, respectively.


We first revisit the BH waterfall configuration \cite{Larre2012}, depicted in Fig. \ref{fig:BHWFConfigurations}a, created by introducing a semi-infinite attractive potential
\begin{equation}\label{eq:waterfall}
V(x)=-V_0\theta(x-x_H),~V_0=V_0(v)=\frac{1}{2}\left(v^2+\frac{1}{v^2}\right)-1
\end{equation}
where $v$ is the asymptotic subsonic Mach number and $x_H$ the point where the step is placed.

The corresponding stationary GP wave function describing the BH configuration is
\begin{widetext}
\begin{equation}\label{eq:CompactBHWF}
\Psi_0(x)=\Psi^{\textrm{BH}}(x)=\left\{ \begin{array}{cc}
e^{ivx}e^{-i\phi_0}\left(v+i\sqrt{1-v^2}\tanh\left[\sqrt{1-v^2}\left(x-x_H\right)\right]\right) & x< x_H\\
ve^{i\frac{x}{v}}, &  x \geq x_H
\end{array}\right.
\end{equation}
\end{widetext}
with $\phi_0$ some phase to make the wave function continuous at $x=x_H$. This configuration provides a realistic theoretical model of the experimental setup of Refs. \cite{Lahav2010,Steinhauer2016,deNova2019}, in which a step potential is swept with velocity $v$ along a condensate at rest, an equivalent situation to that described here via Galilean invariance. We note that this family of BH solutions is uniparametric since it is determined by solely the value of $v$.



As explained in Ref. \cite{deNova2017a}, a stationary BHL solution of arbitrary size can be associated to every stationary BH solution. The argument goes as follows: consider a BH configuration formed by a potential $V^{\textrm{BH}}(x)$ and described by a stationary GP wave function of the form
\begin{equation}\label{eq:BHGP}
\Psi_0(x)=\Psi^{\textrm{BH}}(x)= \left\{ \begin{array}{cc}
    \Psi_u(x)~& x<0\\
    \Psi_d(x)~& x\geq 0\\
\end{array}\right.
\end{equation}
where $\Psi_d(x)=\sqrt{n_d}e^{iv_dx}$ is here the homogeneous plane wave characterizing the supersonic region and we set $x_H=0$ without loss of generality. Then, the configuration formed by the symmetric potential
\begin{equation}\label{eq:potentialsetup}
V^{\textrm{BHL}}(x)= \left\{ \begin{array}{lc}
    V^{\textrm{BH}}\left(\frac{X}{2}+x\right)~& x<0\\
    V^{\textrm{BH}}\left(\frac{X}{2}-x\right)~& x\geq 0\\
\end{array}\right.
\end{equation}
with $V^{\textrm{BHL}}(x)=V^{\textrm{BHL}}(-x)$,
admits as a stationary solution the GP wave function
\begin{widetext}
\begin{equation}\label{eq:BHLsetup}
\Psi_0(x)=\Psi^{\textrm{BHL}}(x)= \left\{ \begin{array}{lr}
    \Psi_{u}\left(\frac{X}{2}+x\right)e^{-i\phi_d}~& x<-X/2 \, .\\
    \Psi_{d}(x)~& -X/2\leq x \leq X/2\\
    \Psi_{u}^*(\frac{X}{2}-x)e^{i\phi_d}~& x>X/2
\end{array}\right.,~\phi_d=\frac{v_d X}{2}
\end{equation}
\end{widetext}
Since $\Psi_0(x)=\Psi^*_0(-x)$, both the sound and flow velocities are symmetric under parity. Therefore, $\Psi_0(x)$ describes a stationary BHL solution as it presents a symmetric pair of horizons, namely, a black hole for $x<0$ and a white hole (WH) for $x>0$. We note that the length $X$ can be chosen arbitrarily, becoming a free parameter.

For the waterfall configuration, such a BHL solution is created from the BH solution of Eq. (\ref{eq:CompactBHWF}) by means of an attractive square well potential of size $X$,
\begin{equation}\label{eq:square}
V(x)=-V_0\theta\left(x+\frac{X}{2}\right)\theta\left(\frac{X}{2}-x\right),
\end{equation}
with the potential depth $V_0$ fine-tuned to the value $V_0=V_0(v)$ given by Eq. (\ref{eq:waterfall}). The resulting BHL configuration is described by the following stationary GP wave function
\begin{widetext}
\begin{equation}\label{eq:CompactBHLWF}
\Psi_0(x)=\Psi^{\textrm{BHL}}(x)=\left\{ \begin{array}{cc}
e^{ivx}e^{-i\phi_0}\left(v+i\sqrt{1-v^2}\tanh\left[\sqrt{1-v^2}\left(x+\frac{X}{2}\right)\right]\right) & x< -\frac{X}{2}\\
ve^{i\frac{x}{v}}, & -\frac{X}{2}\leq x \leq \frac{X}{2}\\
e^{ivx}e^{i\phi_0}\left(v+i\sqrt{1-v^2}\tanh\left[\sqrt{1-v^2}\left(x-\frac{X}{2}\right)\right]\right) & x>\frac{X}{2}
\end{array}\right.
\end{equation}
\end{widetext}
which can be seen in Fig. \ref{fig:BHWFConfigurations}b. This solution is determined by two parameters, $v$ and $X$. Indeed, we can regard this BHL solution as an extension for $X>0$ of the stationary soliton solution when no potential well is present \cite{Langer1967,Sols1994}. The particular relevance of this BHL model is that, while being simple and allowing for an analytical study of a number of its properties due to its homogeneity in the supersonic region, it is similar to the actual configuration observed in the laboratory \cite{Steinhauer2014}, created by sweeping a finite attractive well along a trapped condensate. Moreover, since it is a stationary solution, the pure black-hole lasing contribution can be isolated from other possible background phenomena such as Bogoliubov-Cherenkov radiation \cite{Carusotto2006}, so a study of the genuine black-hole laser properties can be carried out with the perspective of a clear identification in future experiments.

We will refer to Eq. (\ref{eq:CompactBHLWF}) as the \textit{fine-tuned} BHL solution since the value of $V_0$ is fixed by the value of $v$, and the solution inside the well is exactly a plane wave. Other types of BHL solutions, displaying a pair of BH-WH horizons, can be found for the same potential and asymptotic boundary conditions \cite{deNova2017a}. Two main families of BHL solutions can be distinguished: symmetric, $\Psi^S_n(x)=\Psi^{S*}_n(-x)$, and asymmetric, $\Psi^A_n(x)\neq \Psi^{A*}_n(-x)$, with $n=0,1,2\ldots$ The drawback is that they are not homogeneous within the well but instead are described in terms of complicated elliptic functions. Another disadvantage is that each type of solution only exists within a restricted range of values of $X$ for a given value of $v$.

In a general experimental situation, however, $V_0$ is not locked to the value $V_0(v)$. When exploring the complete parameter space $(v,X,V_0)$, allowing now the well amplitude $V_0$ in Eq. (\ref{eq:square}) to be an extra degree of freedom, the fine-tuned BHL solution cease to exist. Still, one can find stationary BHL solutions, given by extensions for arbitrary $V_0$ of the symmetric and asymmetric families of solutions described in the previous paragraph. We will denote these solutions as \textit{generic} BHL solutions. An example of a generic symmetric BHL solution is depicted in Fig. \ref{fig:BHWFConfigurations}c. A more detailed discussion about these families of solutions can be found in Sec. \ref{subsec:NonLinear}.

Finally, the last configuration considered in this work (see Sec. \ref{sec:PhaseDiagram}) is that in which an attractive square well is suddenly introduced in an initially homogeneous flowing condensate (IHFC). In contrast to the previous cases, this setup does not contain horizons and thus is not a BHL setup. An important feature of this model is that the described configuration is not stationary, so it evolves in time deterministically with the homogeneous plane wave $\Psi_0(x)=e^{ivx}$ acting as an initial condition. This configuration is then characterized by the same set of parameters $(v,X,V_0)$ as the generic BHL solutions, providing a useful tool to complement the study of their time evolution. A scheme of this configuration is depicted in Fig. \ref{fig:BHWFConfigurations}d.

\section{Linear and nonlinear spectra}\label{sec:Preview}

Before focusing on the time evolution of the BHL solutions, we first study their linear spectrum of excitations, described by the usual Bogoliubov-de Gennes (BdG) equations, and the full nonlinear spectrum of stationary GP solutions coexisting for the same background parameters and asymptotic boundary conditions. This analysis is expected to help identifying the main features of the time evolution and elaborate some predictions \textit{a priori} \cite{Michel2013,Michel2015,deNova2016}.

\subsection{Linear spectrum}\label{subsec:Linear}

\begin{figure*}[!htb]
\begin{tabular}{@{}lr@{}}
    \includegraphics[width=\columnwidth]{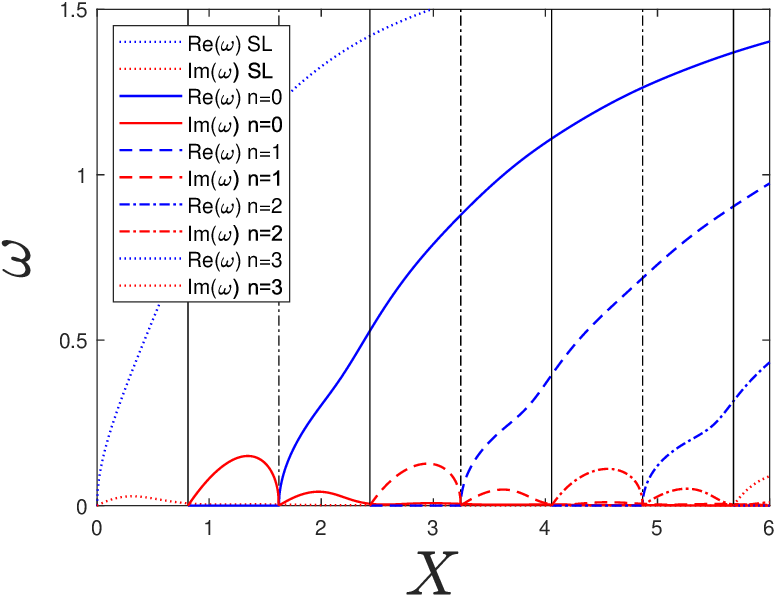} ~~~~~
    & \includegraphics[width=\columnwidth]{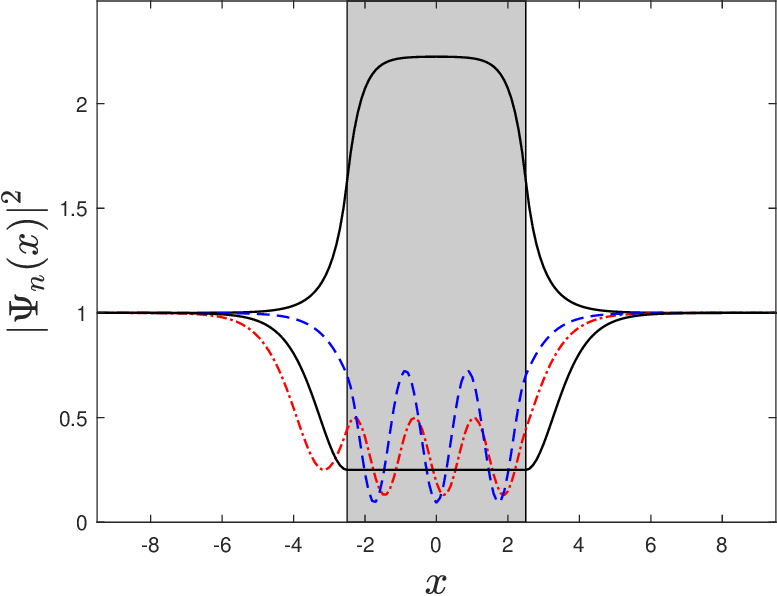} \\
\end{tabular}
\caption{Left panel: complex frequency $\omega$ of the dynamical instabilities, computed from the BdG equations, as a function of the cavity length $X$ for a fine-tuned BHL solution with $v=0.5$. The vertical black lines mark the critical lengths $X^S_n$ (solid) and $X^A_n$ (dash-dotted). Right panel: density profile of some stationary GP solutions in a fine-tuned square well with $v=0.5$ and $X=5$, for which $V_0=V_0(v)=1.125$. Upper solid black: the ground state, $\Psi^{\rm SH}_0(x)$. Lower solid black: the fine-tuned BHL solution, $\Psi^{\rm{BHL}}(x)$. Dashed blue: $n=2$ symmetric solution $\Psi^{S}_2(x)$. Dash-dotted red: $n=2$ asymmetric solution $\Psi^{A}_2(x)$. }
\label{fig:BHLSpectra}
\end{figure*}

We compute in this subsection the linear spectrum of dynamical instabilities as given by the BdG equations. We restrict the analysis to the case of the fine-tuned BHL solution of Eq. (\ref{eq:CompactBHLWF}) since analytical results can be obtained in this background, and it exists for arbitrary cavity length $X$, thus allowing for the use of $X$ as a control parameter to study the behavior of the instabilities. The results for the linear spectrum of this subsection then complement the study of Ref. \cite{Michel2013}, restricted to the idealized flat-profile configuration, and extend the work of Ref. \cite{deNova2017a}, only concerned about the nonlinear spectrum of stationary GP solutions (see next section). The procedure for computing complex-frequency modes of the BdG equations in a BHL is explained in detail in Refs. \cite{Michel2013,Michel2015}.

The spectrum of dynamical instabilities is represented in the left panel of Fig. \ref{fig:BHLSpectra} as a function of the cavity length $X$. We observe that, for the critical lengths
\begin{equation}\label{eq:BHLWFCriticalLengths}
X^S_n=\frac{(2n+1)\pi}{k_0},~k_0=2\sqrt{\frac{1}{v^2}-v^2},~n=0,1,2\ldots
\end{equation}
$k_0$ being the wave vector of the Bogoliubov-Cherenkov mode \cite{Carusotto2006}, a new dynamically unstable mode appears with a purely imaginary frequency (representing a degenerate mode since it is its own BdG conjugate, $\omega=-\omega^*$), but acquiring a non-vanishing real part of the frequency (hence becoming non-degenerate as $\omega\neq-\omega^*$) in the middle of the interval $2\pi/k_0$ between the onsets of instabilities, i.e., at lengths
\begin{equation}\label{eq:BHLWFCriticalLengthsAsymmetric}
X^A_n=X^S_{n+\frac{1}{2}}=\frac{2(n+1)\pi}{k_0},~n=0,1,2\ldots
\end{equation}
The dominant mode (that with the largest growth rate) is typically that with the largest $n$, except close to the critical lengths $X^S_n,X^A_n$. We note the strong similarities with the results for the flat-profile configuration \cite{Michel2013}.

There is however a remarkable difference here: the system is dynamically unstable for any cavity length, since already at $X=0$ a mode with complex frequency and non-vanishing real part emerges. We will refer to this novel mode as the short-length (SL) mode. In contrast, the flat-profile BHL solution lacks the SL mode and is only dynamically unstable for cavity lengths $X>X_0$, with $X_0$ the finite critical length at which the first dynamical instability appears \cite{Michel2013}.

\subsection{Nonlinear spectrum}\label{subsec:NonLinear}

\begin{figure*}[!htb]
\begin{tabular}{@{}cccc@{}}
    \includegraphics[width=0.5\columnwidth]{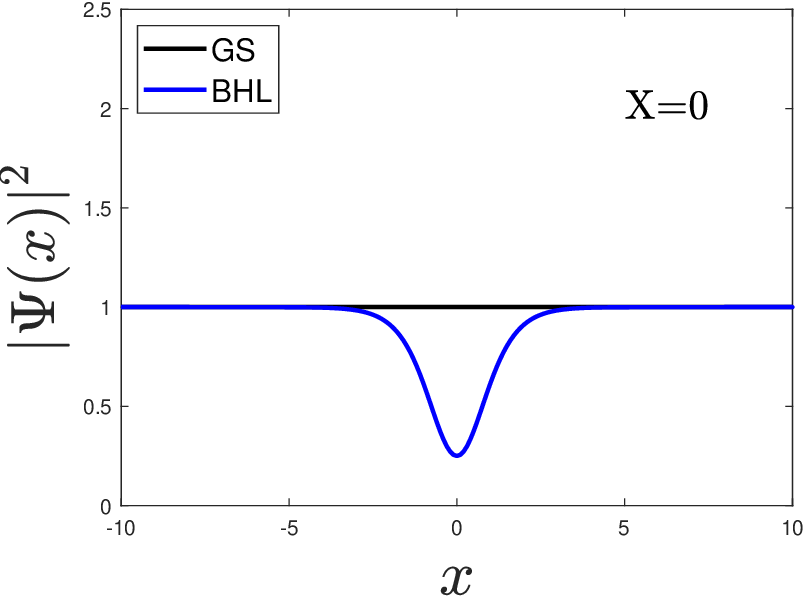} & \includegraphics[width=0.5\columnwidth]{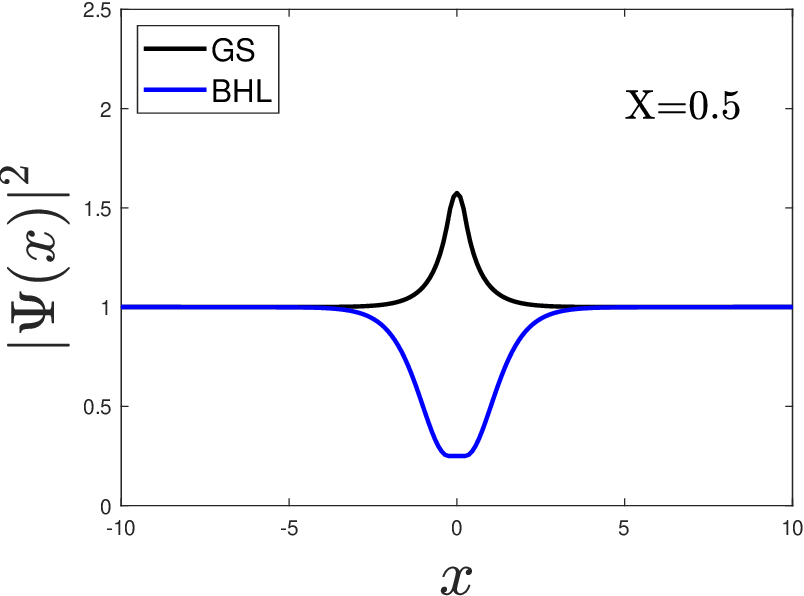} & \includegraphics[width=0.5\columnwidth]{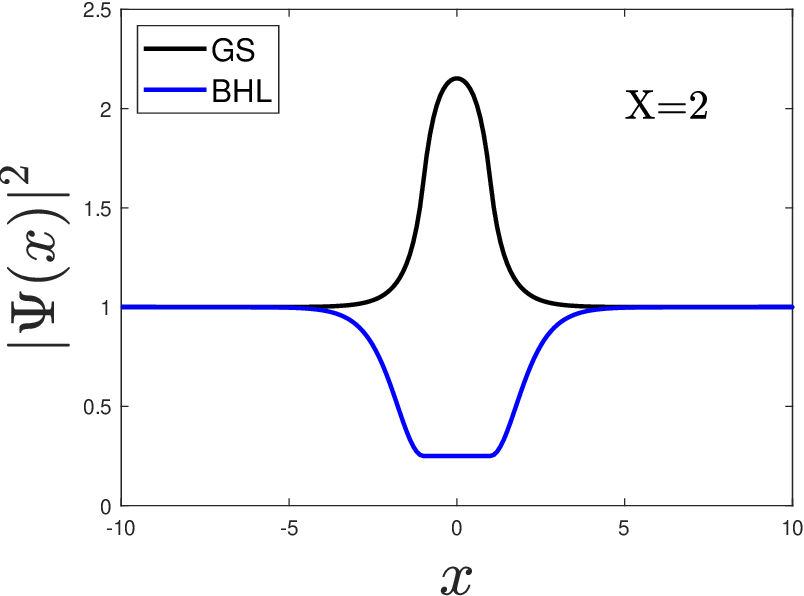} & \includegraphics[width=0.5\columnwidth]{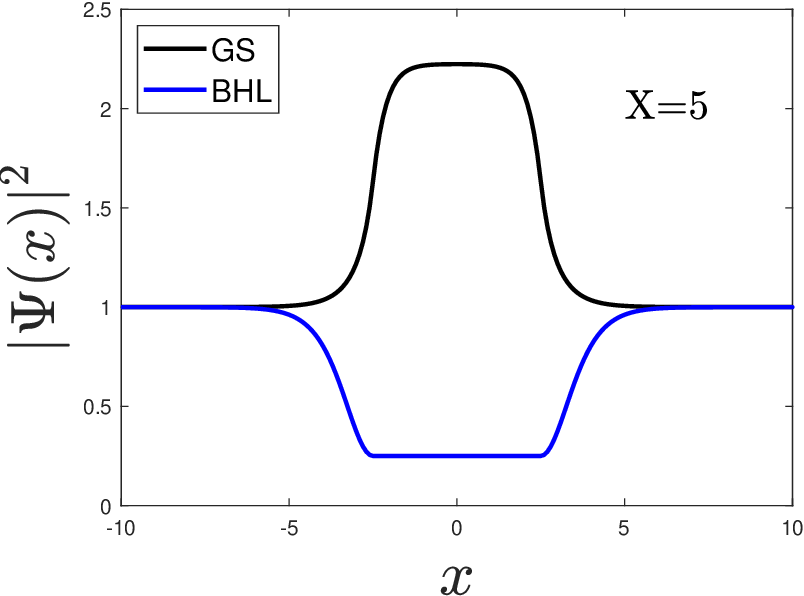} \\ [0.5cm]
    \includegraphics[width=0.5\columnwidth]{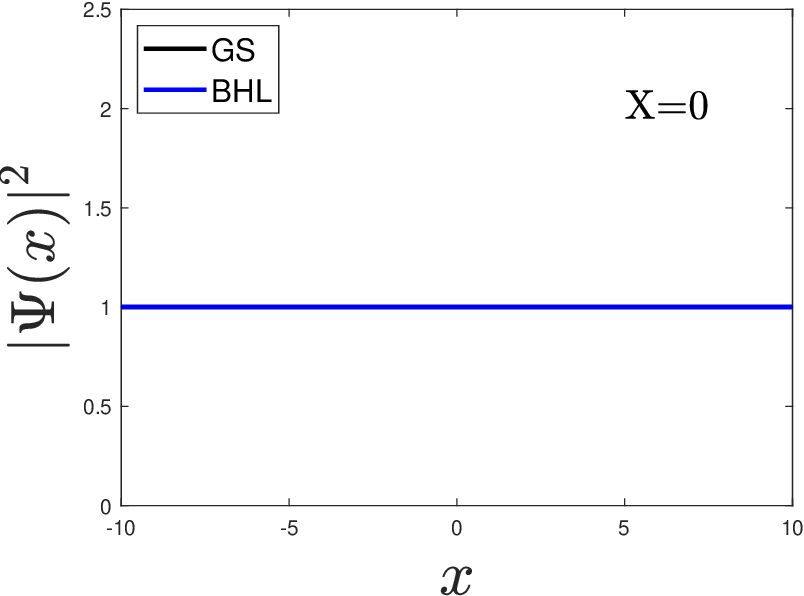} & \includegraphics[width=0.5\columnwidth]{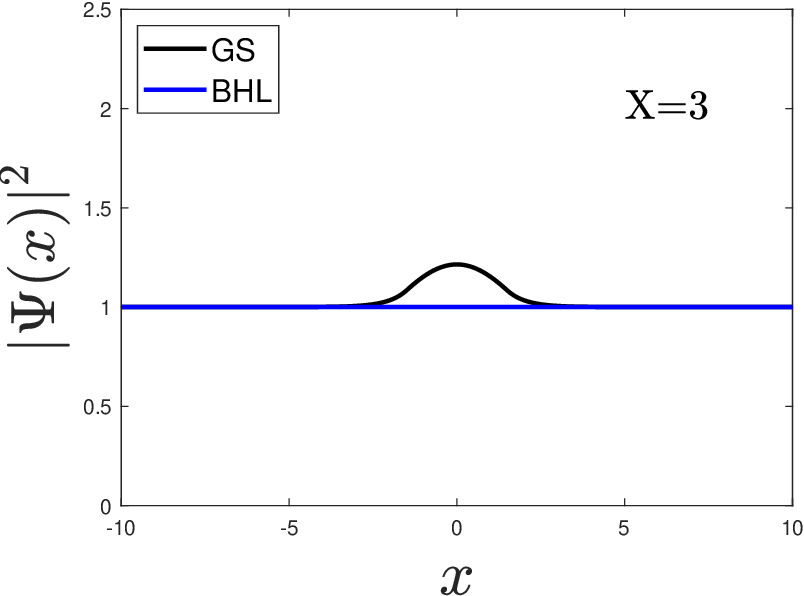} & \includegraphics[width=0.5\columnwidth]{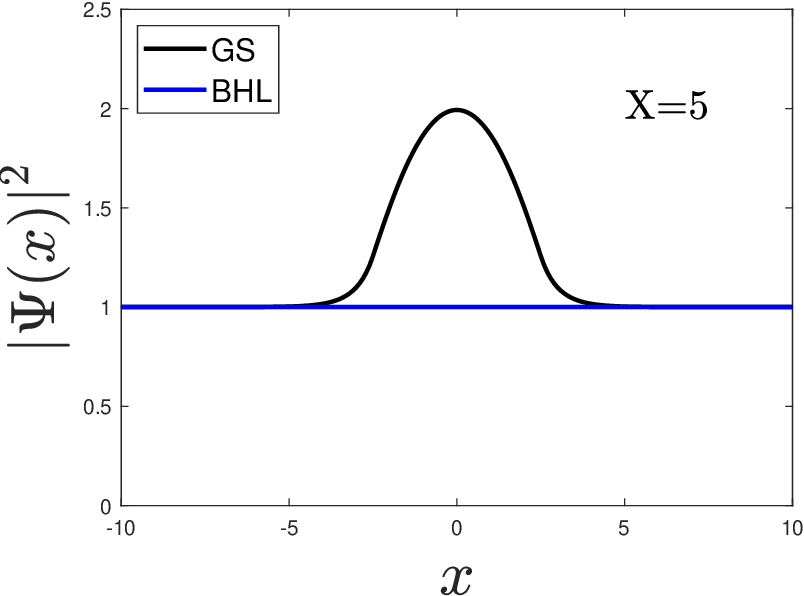} & \includegraphics[width=0.5\columnwidth]{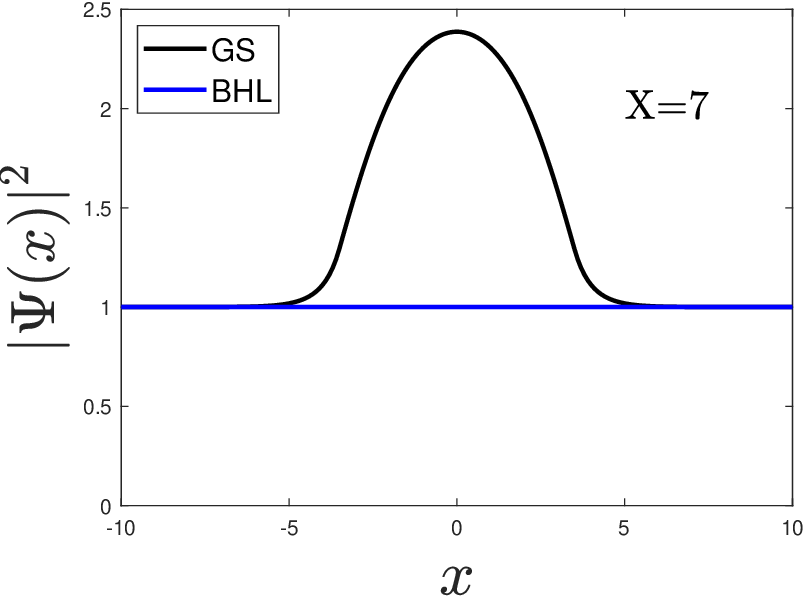} \\
\end{tabular}
\caption{Upper row: density profile of the BHL (blue) and GS (black) solutions for increasing values of $X$ for a fine-tuned configuration with $v=0.5$. Lower row: same as upper but for a flat-profile configuration \cite{Michel2013} with $v=0.5$ and $c_2=0.25$, with critical length $X_0=2.56$.}
\label{fig:BHLTopology}
\end{figure*}

The results for the linear BdG spectrum can be understood in terms of the nonlinear spectrum of stationary GP solutions. A detailed study of the full nonlinear GP spectrum in a fine-tuned square well with $V_0=V_0(v)$ was provided in Ref. \cite{deNova2017a}. We briefly revisit here the results of that work and connect them with the computation of the linear BdG spectrum of instabilities performed in the previous subsection. We also extend those results to arbitrary (non fine-tuned) values of the well amplitude $V_0$.

The most relevant families of solutions for the present matter are three: symmetric BHL, asymmetric BHL (both already introduced in Sec. \ref{sec:WF}) and the ground-state family. The first two families of solutions emerge precisely at critical lengths $X^S_n,X^A_n$, described in this limit by small-amplitude oscillations on top of the fine-tuned BHL solution. Thus, these families of solutions are smoothly connected to the fine-tuned BHL solution as a function of $X$, which explains important features of the BdG spectrum: the appearance of the symmetric $\Psi^{S}_n(x)$ solution at $X=X^S_n$ describes the appearance of the degenerate $n$-th lasing mode while the appearance of the asymmetric $\Psi^{A}_n(x)$ solution at $X=X^A_n$ describes the transformation of the $n$-th mode into a non-degenerate BdG mode, in the same fashion as for the flat-profile case \cite{Michel2013,Michel2015}. In turn, the symmetric family can be divided into complete and incomplete solutions, depending on whether the soliton outside the well has reached its minimum or not, respectively. The incomplete solutions are more energetically favorable so in the following we will generally refer to the symmetric incomplete BHL solutions as simply symmetric BHL solutions, unless otherwise stated. The density profile for a symmetric and asymmetric BHL solution is depicted in the right panel of Fig. \ref{fig:BHLSpectra}, along with that for the fine-tuned BHL solution.

The remaining relevant family of solutions is labeled as $\Psi^{\rm SH}_n(x)$ because they are described in terms of so-called shadow solitons \cite{Michel2013} outside the cavity, instead of regular solitons as the previous BHL families. The wave functions $\Psi^{\rm SH}_n(x)$ are symmetric, with $n=0,1,2\ldots$ the number of periods inside the well. The importance of this family of solutions is that the $n=0$ solution describes the \textit{true} ground state of the system, representing the lowest-energy GP solution compatible with the given asymptotic boundary conditions. The stability of $\Psi^{\rm SH}_0(x)$ results from the accumulation of atoms in the central region in order to become fully subsonic. By means of this process the system clears away all the horizons, in what resembles the evaporation of a real black hole. The density profile of the ground state is represented in solid black line in the right panel of Fig. \ref{fig:BHLSpectra}.

The ground state already emerges at $X=0$, departing smoothly from the homogeneous subsonic plane wave, which is the ground state when no potential well is present. Thus, as a function of $X$, the ground state $\Psi^{\rm SH}_0(x)$ is smoothly connected to the homogeneous plane wave solution, while the fine-tuned BHL solution is smoothly connected to the soliton solution [see Eq. (\ref{eq:CompactBHLWF}) and ensuing discussion]. This smooth departure is observed in the upper row of Fig. \ref{fig:BHLTopology}, where both the ground state and the fine-tuned BHL solution are represented for increasing cavity length. Similar scenarios in which nonlinear stationary solutions smoothly depart from the homogeneous and soliton solutions as an external potential is introduced were already studied in the context of the crossover from the Josephson effect to bulk superconducting flow within a Ginzburg-Landau description \cite{Sols1994}.

Therefore, the fine-tuned BHL solution is never the true ground state of the system. This contrasts with the case of the flat-profile configuration, represented in the lower row of Fig. \ref{fig:BHLTopology}, in which the homogeneous BHL solution is still the ground state for cavity lengths below the critical length $X=X_0$. Above that length, the true ground state begins to smoothly depart as a function of $X$ from the BHL homogeneous solution, which becomes dynamically unstable \cite{Michel2013}. Returning to the fine-tuned BHL solution, its excited-state character, along with the conjecture that only the ground state is dynamically stable \cite{Michel2013,Michel2015}, explains the appearance of the SL mode found in the previous subsection.

The different stationary nonlinear solutions also play a key role in the time evolution of the system, since they represent metastable states intercepted during the transient dynamics in which the system stays for some time before collapsing due to their unstable character \cite{Michel2015,deNova2016}. Eventually, only the ground state is expected to be dynamically stable, and hence it is one of the natural ending points of the dynamics.

In the case of arbitrary $V_0$, the situation is similar because the same families characterize the relevant set of stationary nonlinear GP solutions. The main difference with respect to the fine-tuned scenario is that the fine-tuned BHL solution no longer exists and we only have the symmetric and asymmetric configurations available as BHL solutions. An important limitation is that each BHL solution $\Psi^{S}_n(x),\Psi^{A}_n(x)$ only exist in a restricted region of the parameter space $(v,X,V_0)$. As a result, there are sets of values of $(v,X,V_0)$ for which no initial BHL solutions can be found. The family of solutions $\Psi^{\rm SH}_n(x)$ is also present in the generic case. In particular, the ground state solution $\Psi^{\rm SH}_0(x)$ also appears for any nonzero length and, as a function of $X$, is smoothly connected to the homogeneous subsonic solution for $X=0$. A complete study of the full spectrum of stationary GP solutions in a square well potential can be found in Ref. \cite{Leboeuf2001}.

\begin{figure*}[!htb]
\begin{tabular}{@{}cccc@{}}
    \includegraphics[width=0.5\columnwidth]{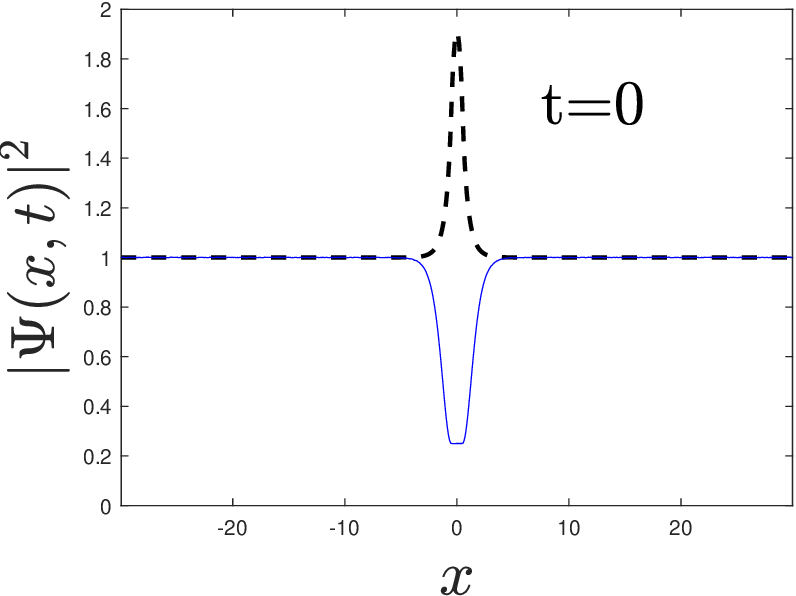} & \includegraphics[width=0.5\columnwidth]{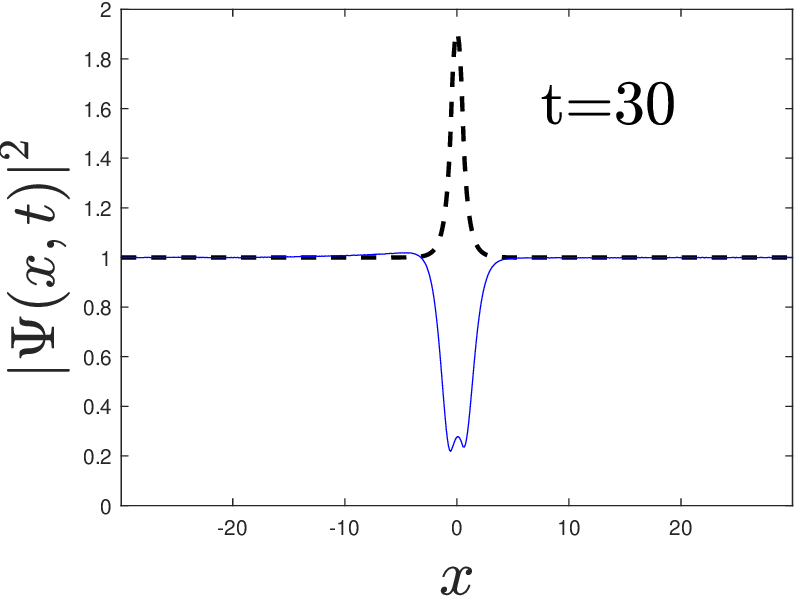} &
    \includegraphics[width=0.5\columnwidth]{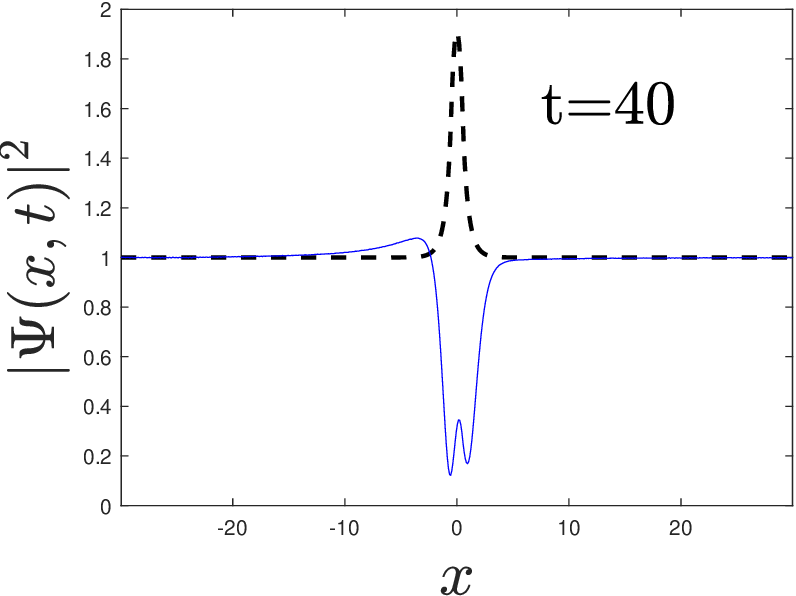} & \includegraphics[width=0.5\columnwidth]{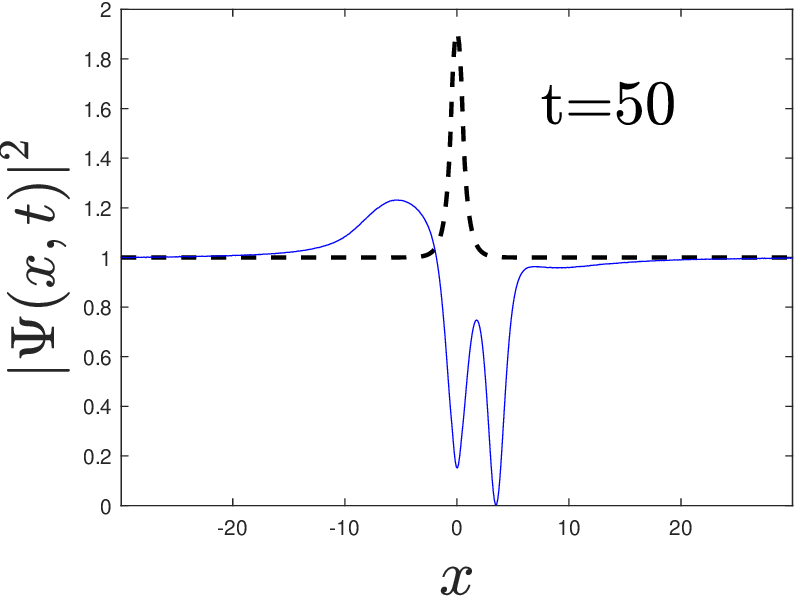} \\[0.5cm]
    \includegraphics[width=0.5\columnwidth]{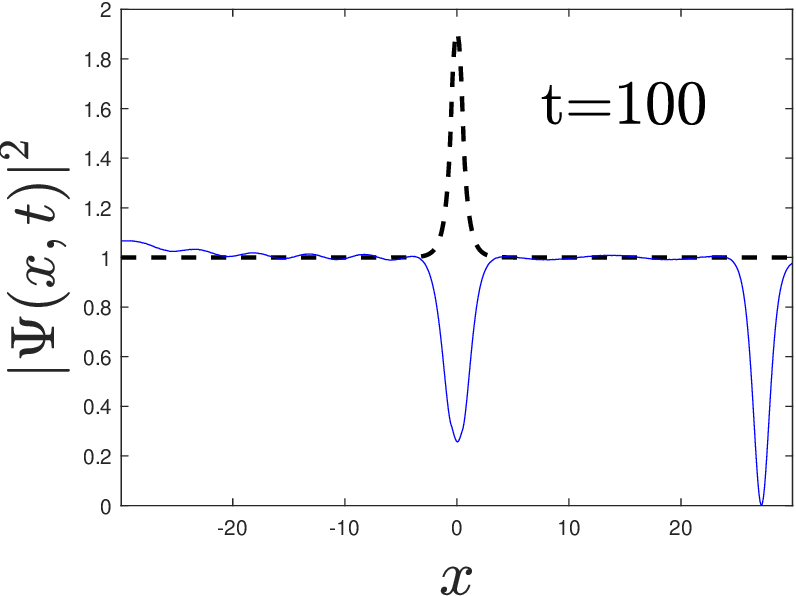} & \includegraphics[width=0.5\columnwidth]{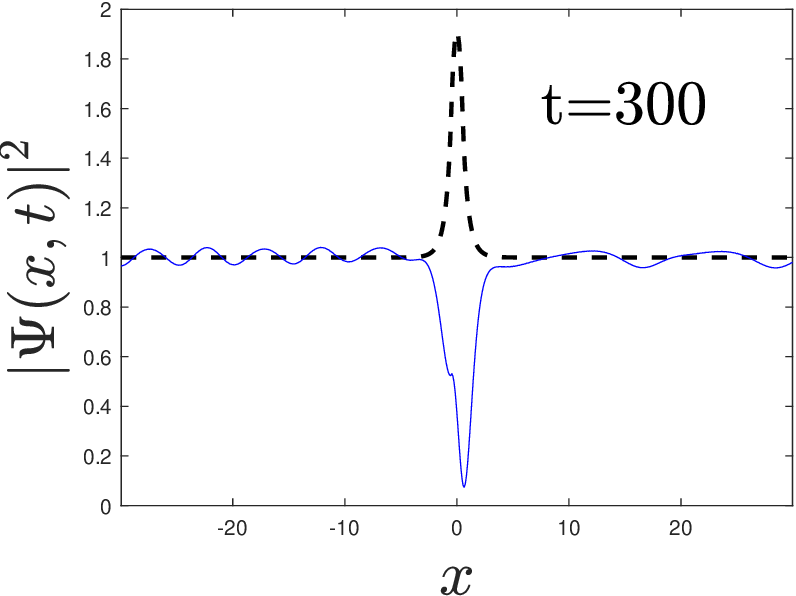} &
    \includegraphics[width=0.5\columnwidth]{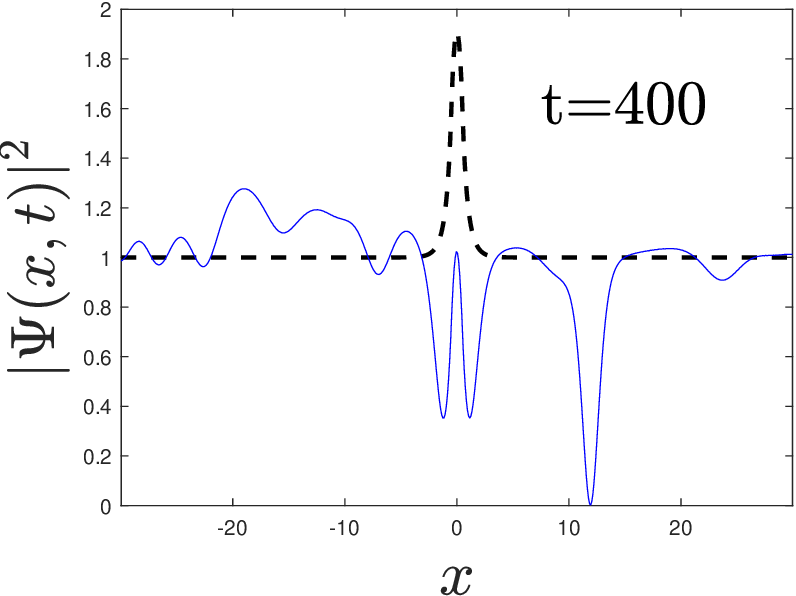} & \includegraphics[width=0.5\columnwidth]{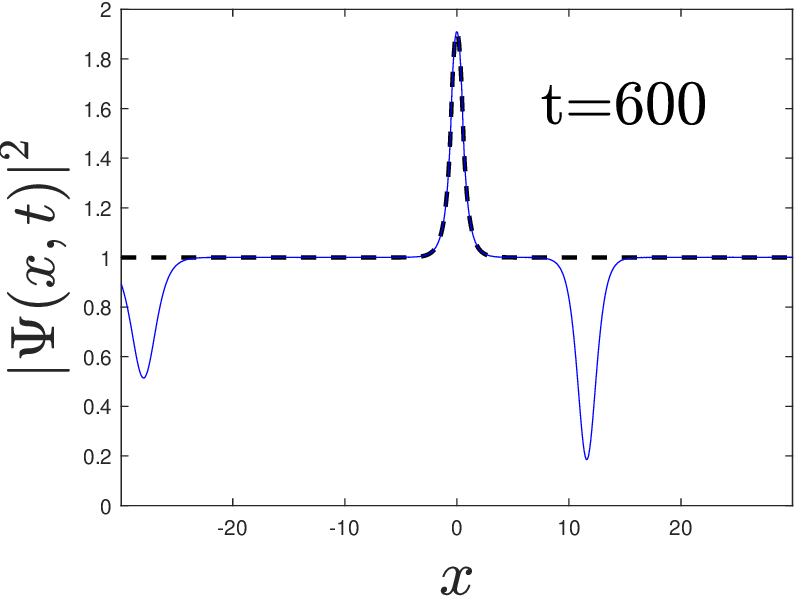} \\[0.5cm]
    \includegraphics[width=0.5\columnwidth]{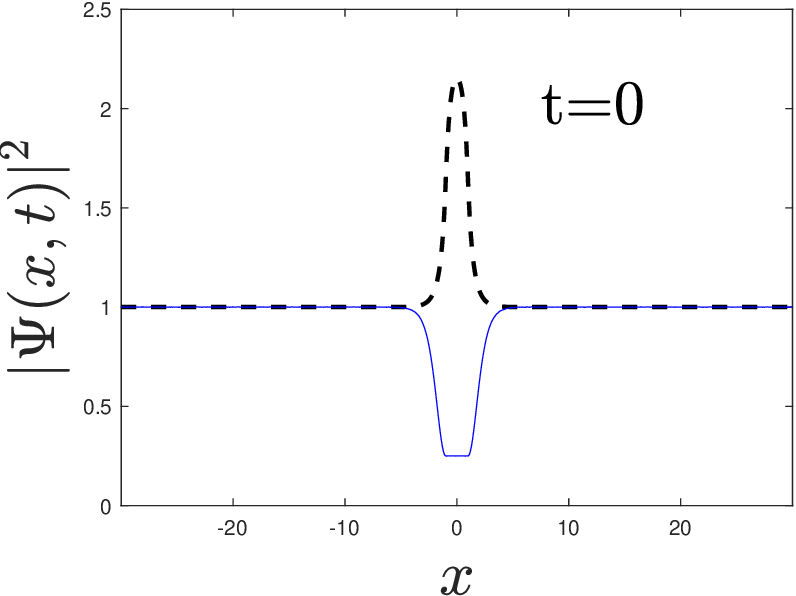} & \includegraphics[width=0.5\columnwidth]{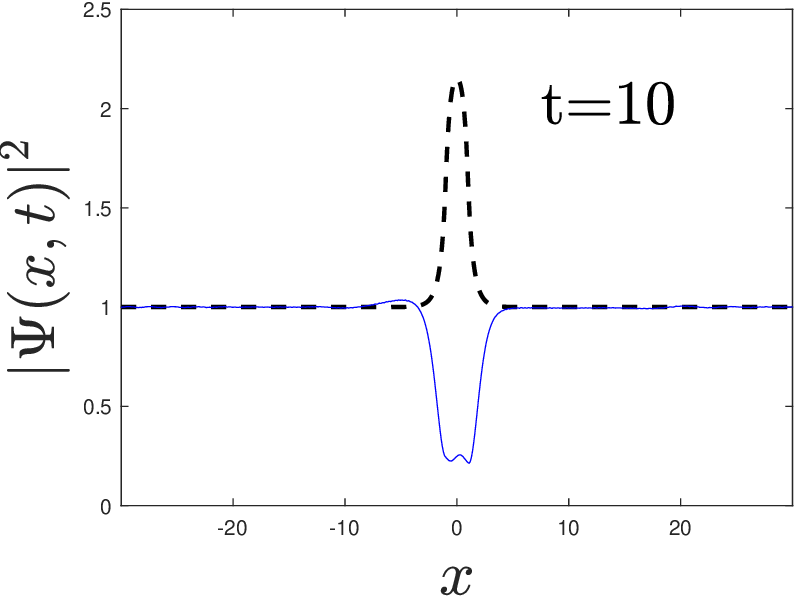} &
    \includegraphics[width=0.5\columnwidth]{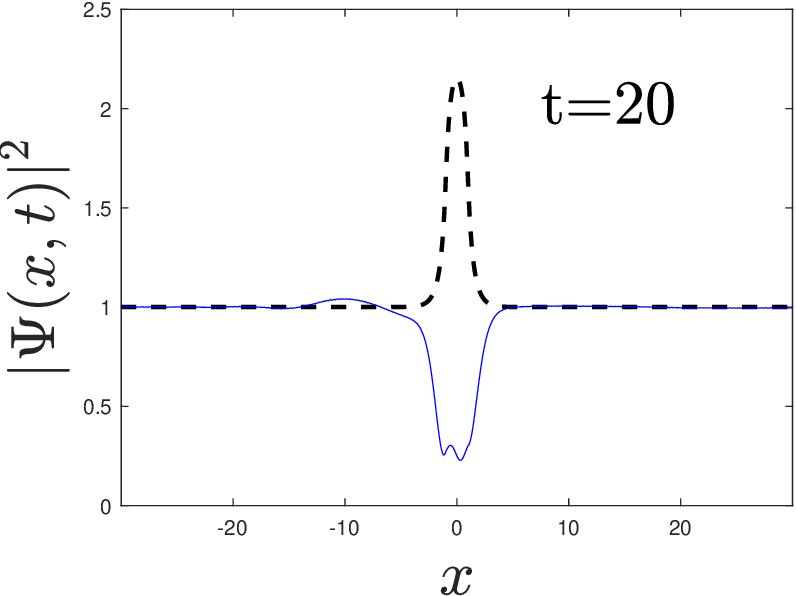} & \includegraphics[width=0.5\columnwidth]{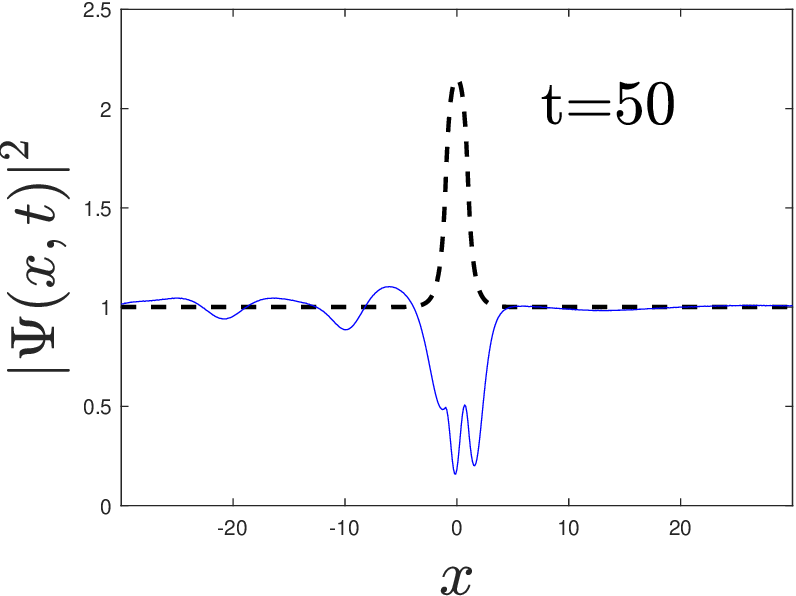} \\[0.5cm]
    \includegraphics[width=0.5\columnwidth]{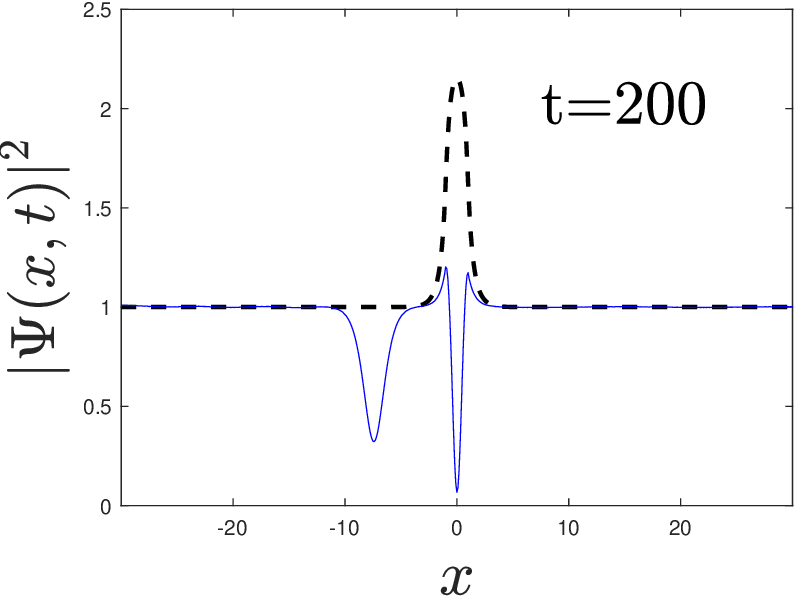} & \includegraphics[width=0.5\columnwidth]{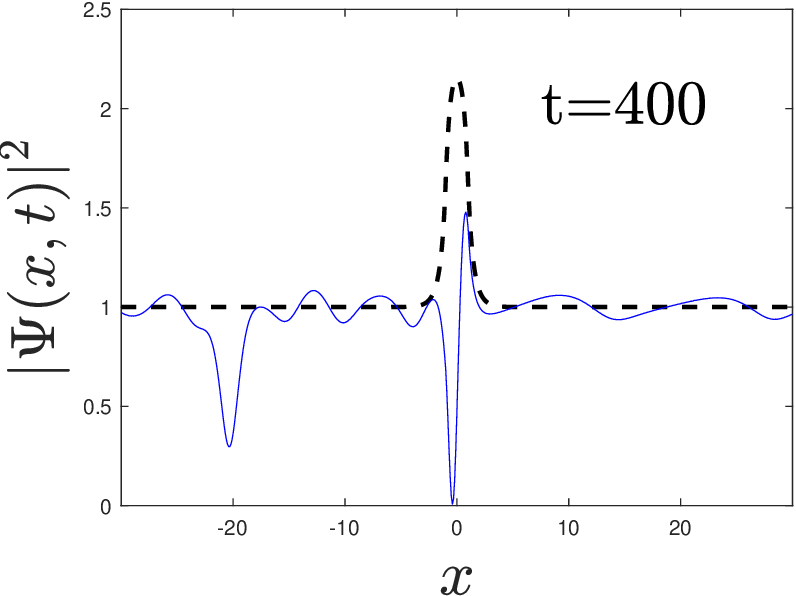} &
    \includegraphics[width=0.5\columnwidth]{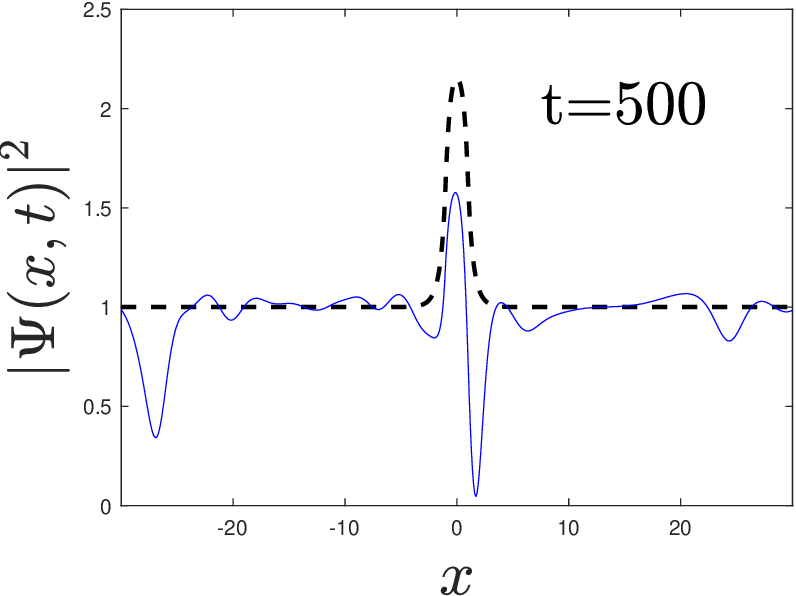} & \includegraphics[width=0.5\columnwidth]{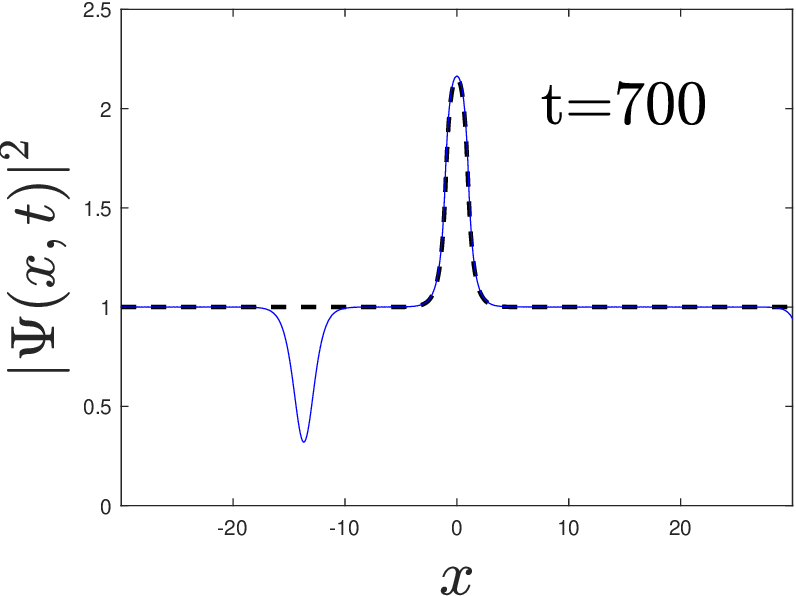}
\end{tabular}
\caption{Plot of the time evolution of the density profile $|\Psi(x,t)|^2$ (solid blue) for different times for an initial fine-tuned BHL. The ground state is depicted in dashed black line. First and second row: initial fine-tuned BHL solution with $v=0.5$ and $X=1$, so that $X^S_0<X<X^A_0$.
Third and fourth row: same but for $X=2$, so that $X^A_0<X<X^S_1$.}
\label{fig:BHLFineTuneTime}
\end{figure*}

\section{Time evolution of realistic black-hole laser setups}\label{sec:TimeEvolution}

In order to numerically study the time evolution of the BHL solutions described in Sec. \ref{sec:WF}, we follow the procedure of Ref. \cite{deNova2016}: we add some small random noise to the stationary BHL solutions in order to trigger the dynamical instabilities, and then let that stochastic initial condition evolve with the time-dependent GP equation.

\subsection{Short-time behavior}\label{subsec:ShortTime}

We first analyze the short times of the simulations, in which the instability has not yet grown enough to enter the full nonlinear regime and thus the dynamics can be still described within the BdG framework. We numerically find that all the BHL solutions are dynamically unstable for any nonzero cavity length. In particular, for the fine-tuned BHL configuration, we compare the numerical results with the predictions from the linear BdG spectrum computed in Sec. \ref{subsec:Linear}, observing a perfect agreement.

More specifically, and as expected from Fig. \ref{fig:BHLSpectra}a, the evolution of the system is typically dominated by the largest $n$-th mode, appearing at the predicted critical lengths $X^S_n$ with purely imaginary frequency, so the observed growth of the resulting density pattern is monotonous in time. An example of this behavior is displayed in the first row of Fig. \ref{fig:BHLFineTuneTime}, where the monotonous growth of the $n=0$ mode can be observed.

However, at the critical lengths $X^A_n=X^S_{n+\frac{1}{2}}$, the frequency of the $n$-th mode acquires a nonzero real part of the frequency and the unstable density pattern grows while oscillating. An example of this behavior is displayed in the third row of Fig. \ref{fig:BHLFineTuneTime}, where the oscillating growth of the $n=0$ mode can be observed.

\subsection{Long-time behavior}\label{subsec:LongTime}

Once the instability, induced by the noise initially added to the BHL solution, has grown enough as a result of the time evolution, the system enters the nonlinear regime. The trend here is that the condensate approaches the different possible stationary GP solutions described in Sec. \ref{subsec:NonLinear} while emitting sound waves and solitons in order to conserve energy and particle number. Nevertheless, we numerically find that all the stationary GP solutions are dynamically unstable except the ground state, so they eventually collapse and the dynamics is restarted again. This confirms the conjecture of Refs. \cite{Michel2013,Michel2015}, which states that within this kind of flowing configurations there is a perfect correspondence between energetic and dynamical instability: each GP solution that is not a strict local minimum of the Hamiltonian (energetic instability) presents unstable complex-frequency BdG modes (dynamical instability). In general, energetic instability is only a necessary condition for dynamical instability, but not sufficient \cite{Wu2003}.

The above described dynamics is continued until sufficiently long times are explored, when only two possible behaviors are observed: the system either reaches GS by evaporating away the horizons or evolves to the CES regime, where it self-oscillates in a time-periodic way. The results of this work then confirm the trends described in Ref. \cite{deNova2016}.

\begin{figure*}[!t]
\begin{tabular}{@{}cccc@{}}
    \includegraphics[width=0.5\columnwidth]{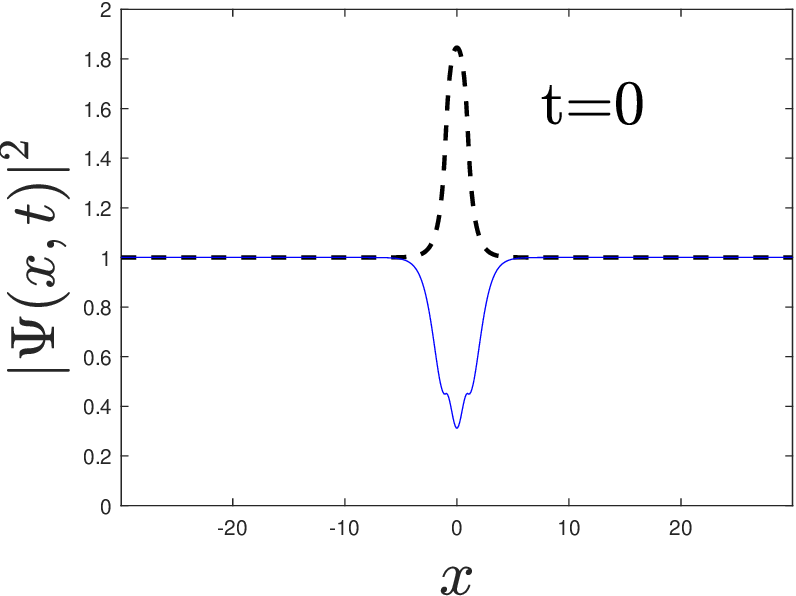} & \includegraphics[width=0.5\columnwidth]{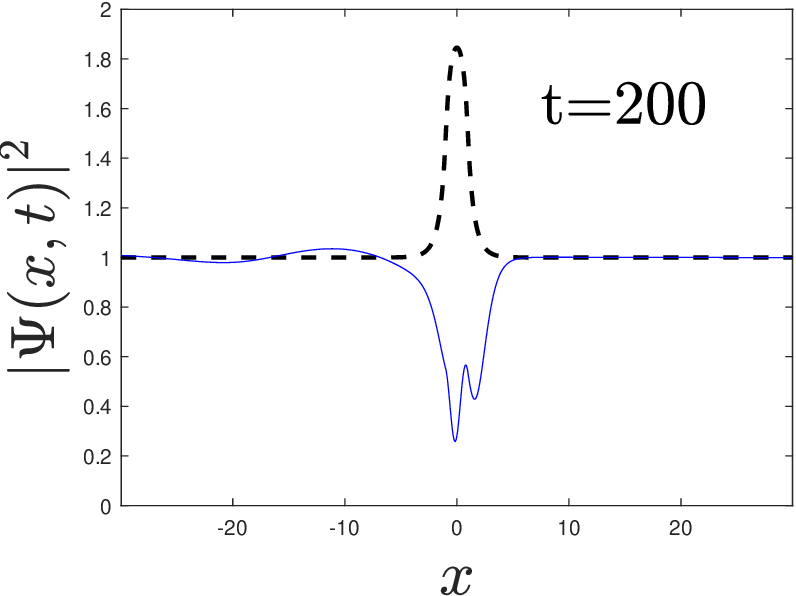} &
    \includegraphics[width=0.5\columnwidth]{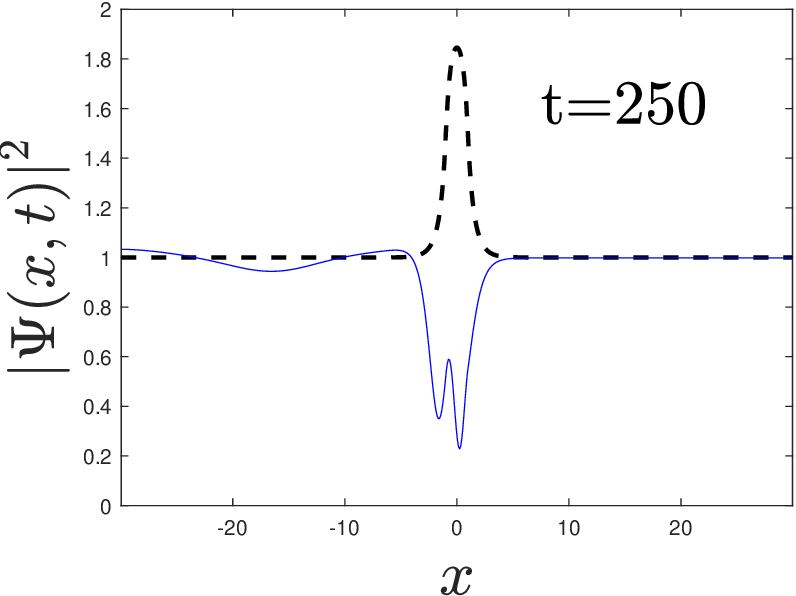} & \includegraphics[width=0.5\columnwidth]{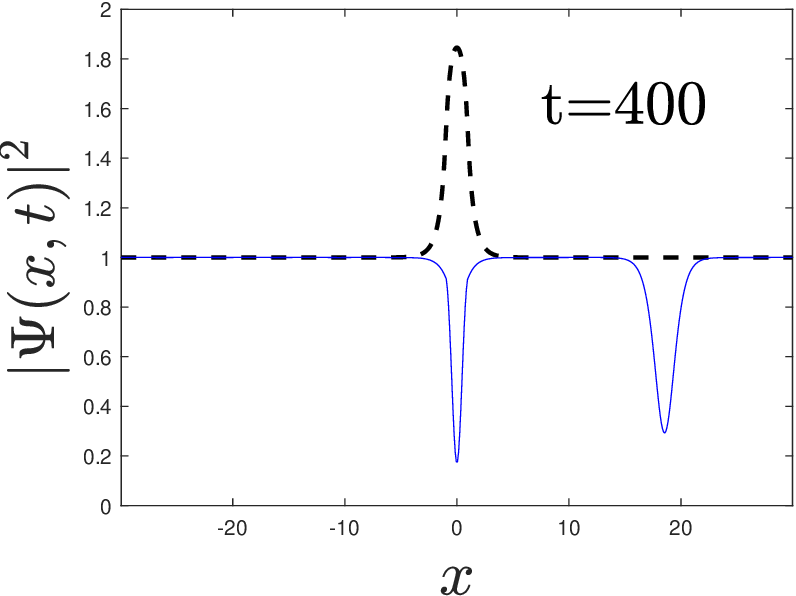} \\[0.5cm]
    \includegraphics[width=0.5\columnwidth]{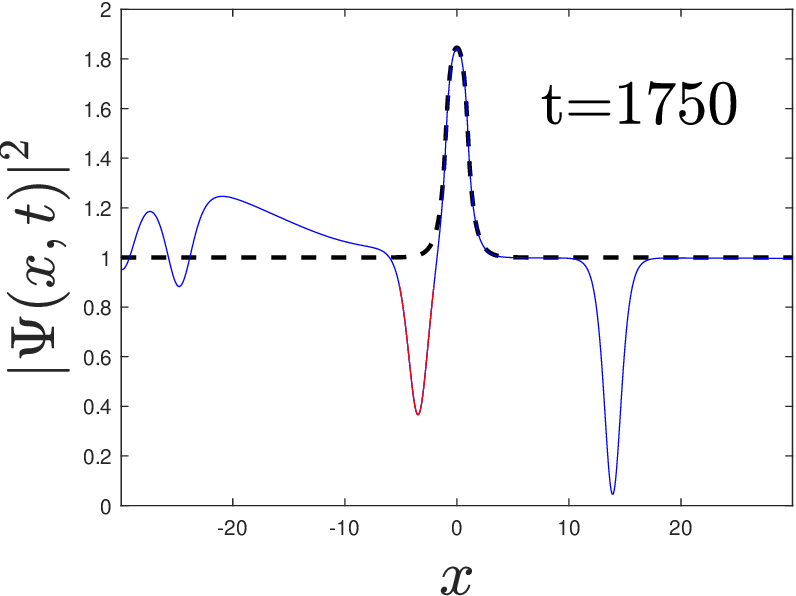} & \includegraphics[width=0.5\columnwidth]{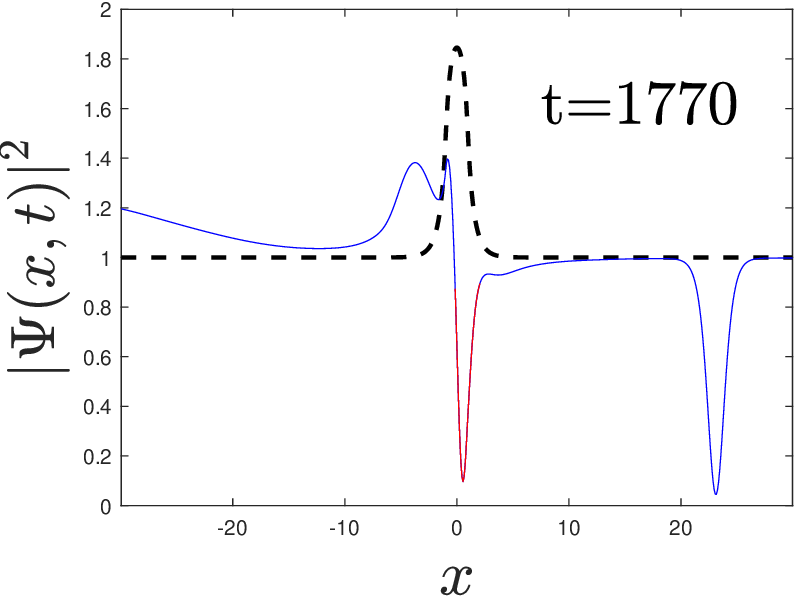} &
    \includegraphics[width=0.5\columnwidth]{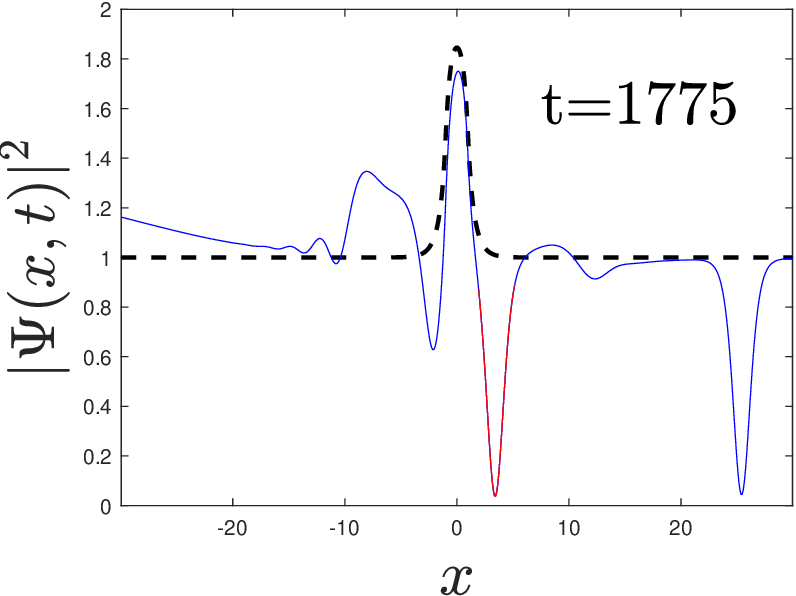} & \includegraphics[width=0.5\columnwidth]{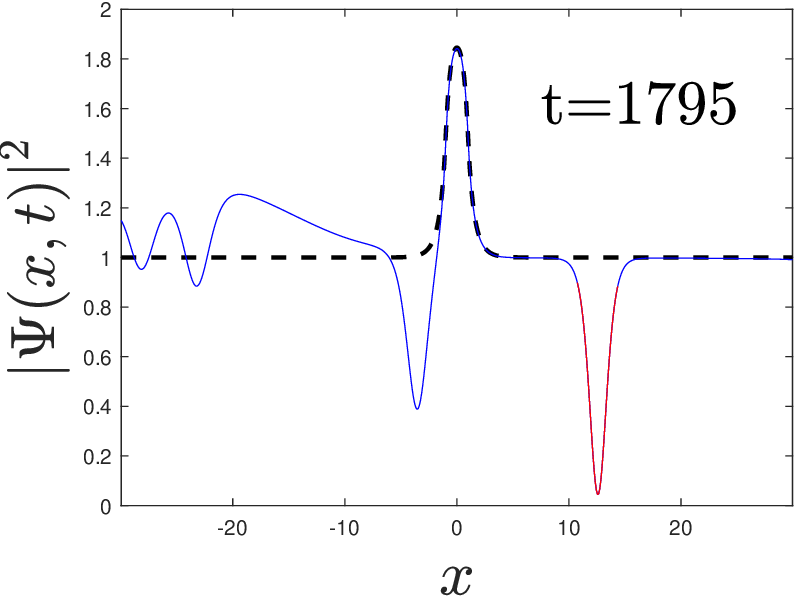}
\end{tabular}
\caption{Same as Fig. \ref{fig:BHLFineTuneTime} but for an initial generic symmetric \textit{complete} BHL $n=0$ solution with $v=0.67$, $V_0=0.75$ and $X=2$. The red color in the lower row helps to identify the soliton whose evolution gives rise to the CES mechanism.}
\label{fig:BHLGenericTime}
\end{figure*}

Representative examples of this dynamics can be seen in Fig. \ref{fig:BHLFineTuneTime}, where we examine the time evolution of fine-tuned BHL solutions. In the first two rows, we observe the monotonous growth of an initially unstable fine-tuned BHL solution towards the $\Psi^S_0(x)$ stationary solution (see $t=100$) while emitting a soliton downstream. However, such solution is also dynamically unstable and eventually collapses, finally reaching the stable ground state while emitting solitons and waves. In the last two rows, the results from another simulation are shown, in which the system grows while oscillating towards the stationary solution $\Psi^{\rm SH}_1(x)$ (see $t=200$), which in turn is also unstable, collapsing for sufficiently long times and reaching once more the stable ground state after radiating. For longer cavities, the evolution becomes much more complicated since more nonlinear solutions are available to be intercepted during the transient. Nevertheless, for sufficiently long times, we have numerically observed that all fine-tuned BHL solutions always evolve towards GS.

In contrast, generic BHL solutions can indeed reach the CES regime, as shown in Fig. \ref{fig:BHLGenericTime}. In the first row, we observe an evolution along the same lines of Fig. \ref{fig:BHLFineTuneTime}: an initially unstable symmetric \textit{complete} $n=0$ BHL solution grows while oscillating towards $\Psi^S_0(x)$ (see $t=400$), emitting a soliton downstream in the process. Such stationary solution is again unstable and the system eventually collapses approaching the ground state. However, as seen in the lower row of Fig. \ref{fig:BHLGenericTime}, the soliton emitted upstream (marked in red; see also the latest times in Fig. \ref{fig:BHLFineTuneTime}) cannot travel now against the flow and is dragged back towards the supersonic region. The passage of the soliton through the square well leaves the system again in the same initial configuration that emits another soliton upstream, giving rise in this way to the periodic self-oscillating process leading to the CES state. The described mechanism is quite similar to that of the flat-profile configuration, characterized in detail in Ref. \cite{deNova2016}.

Even though the initial seed is stochastic, leading to very different possible transients between simulations, the specific long-time fate of the system (GS or CES) is solely determined by the background parameters $(v,X,V_0)$ and not by the initial noise or the type (asymmetric-symmetric or the integer $n$) of initial BHL solution. Only very close to the phase transition from GS to CES some sensitivity to the initial condition can be sometimes observed. We attribute this sensitivity to the critical effect of fluctuations on the amplitude of the emitted soliton that originates the CES state.

This independence of the final state with respect to the transient implies that the equivalent of a phase diagram for the long-time behavior, characterized by the GS and CES states, can be drawn in the parameter space $(v,X,V_0)$: at long times, for fixed values of $(V_0,X)$, there is a critical velocity $v_c\left(V_0,X\right)$ above which the system enters the CES regime.  Specifically, if the flow velocity $v$ is below the critical velocity $v_c\left(V_0,X\right)$, the soliton emitted upstream is able to escape and the system evolves towards the GS. Above that critical velocity, the soliton is trapped by the flow, giving rise to the CES state.

However, the numerical computation of the long-time phase diagram for generic BHL solutions presents several problems. First, there are regions in parameter space in which BHL solutions do not exist. Second, for large cavities, the high number of intermediate metastable states hit by the dynamics, along with the associated decay process, make the simulations computationally expensive due to the long times required. We develop in the next section an alternative approach to address the problem.

\begin{figure*}[!htb]
\begin{tabular}{@{}cccc@{}}
    \includegraphics[width=0.5\columnwidth]{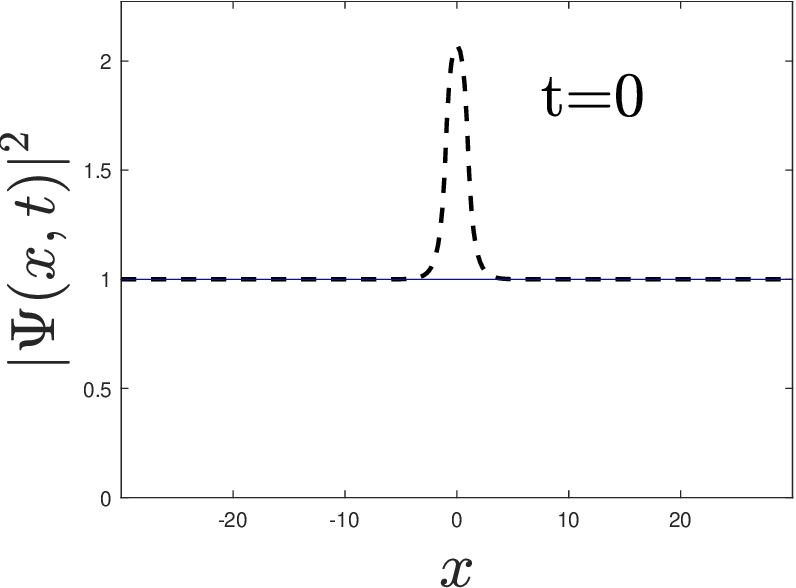} & \includegraphics[width=0.5\columnwidth]{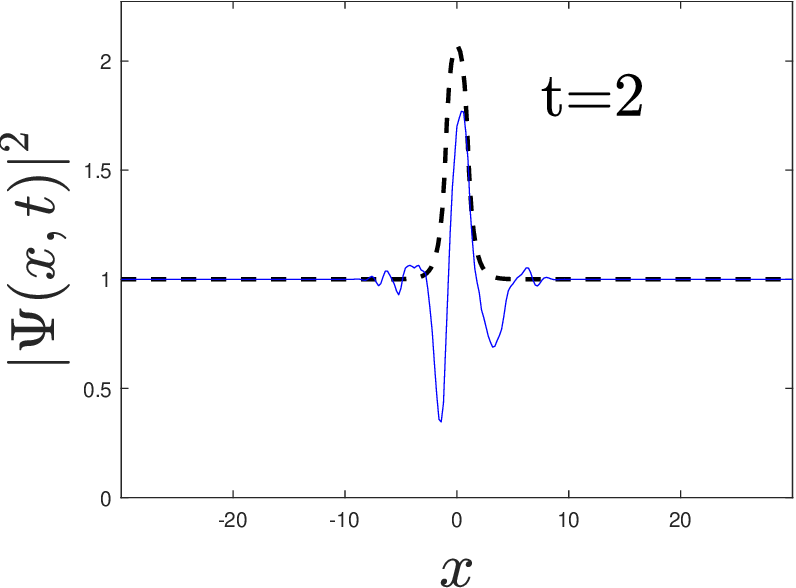} &
    \includegraphics[width=0.5\columnwidth]{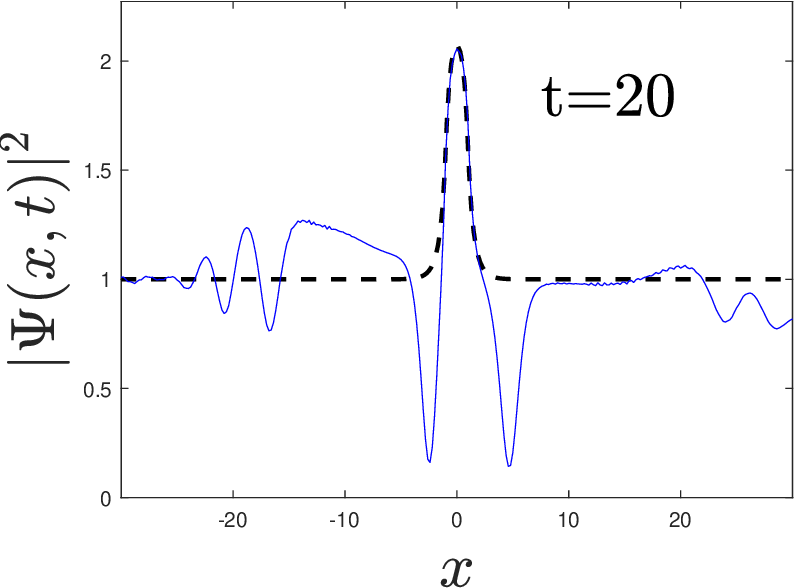} & \includegraphics[width=0.5\columnwidth]{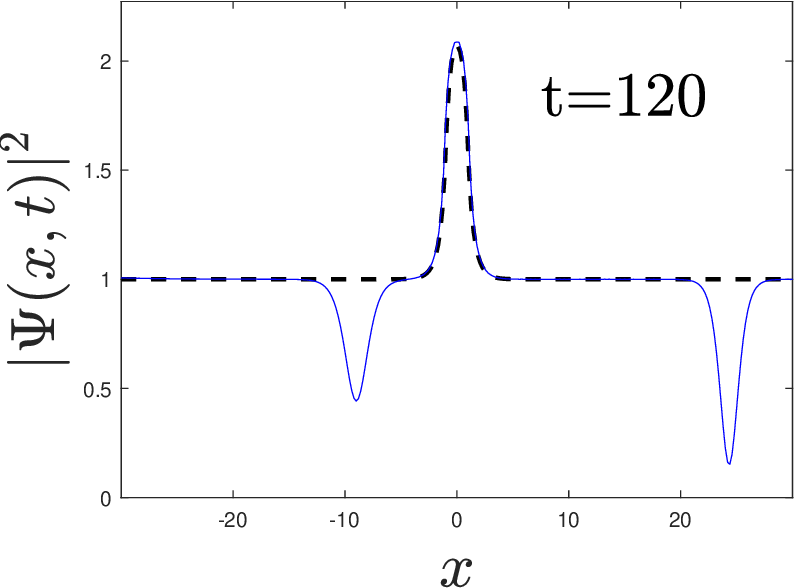} \\[0.5cm]
    \includegraphics[width=0.5\columnwidth]{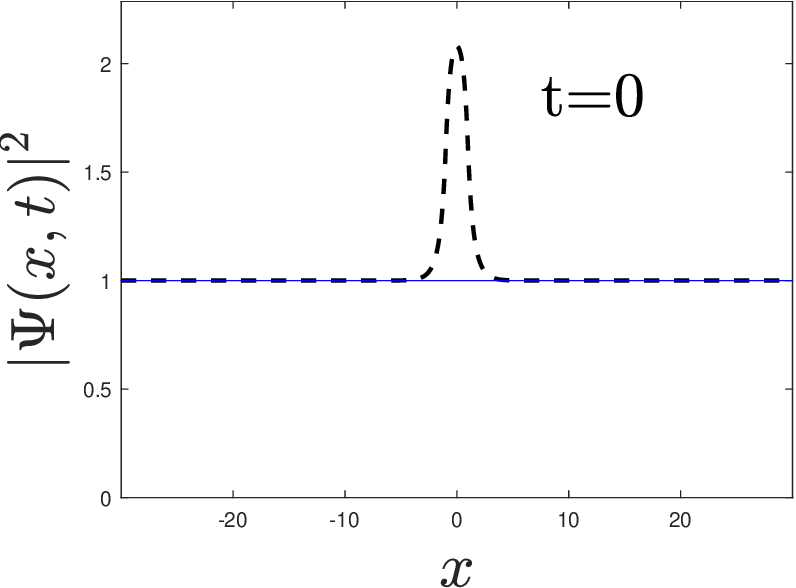} & \includegraphics[width=0.5\columnwidth]{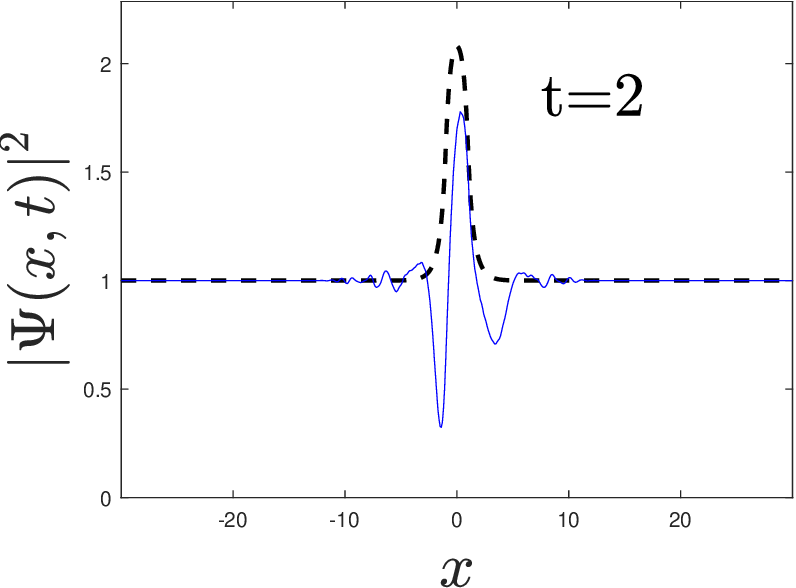} &
    \includegraphics[width=0.5\columnwidth]{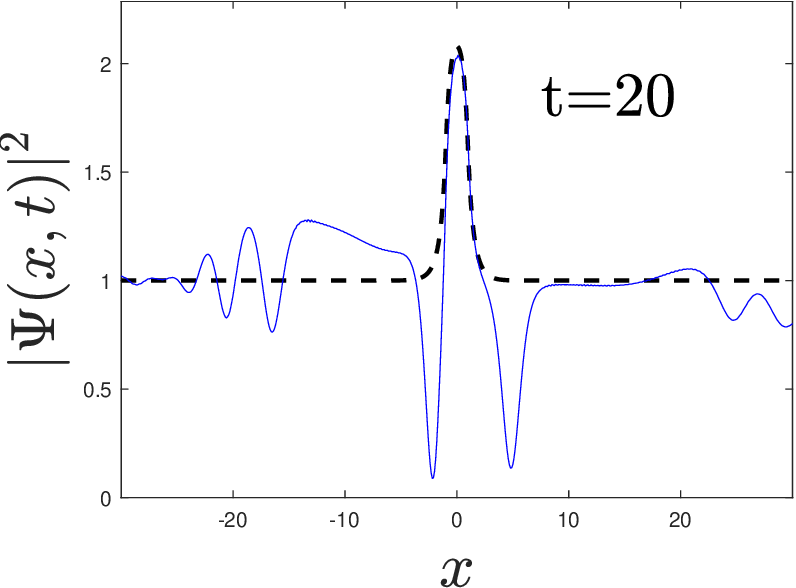} & \includegraphics[width=0.5\columnwidth]{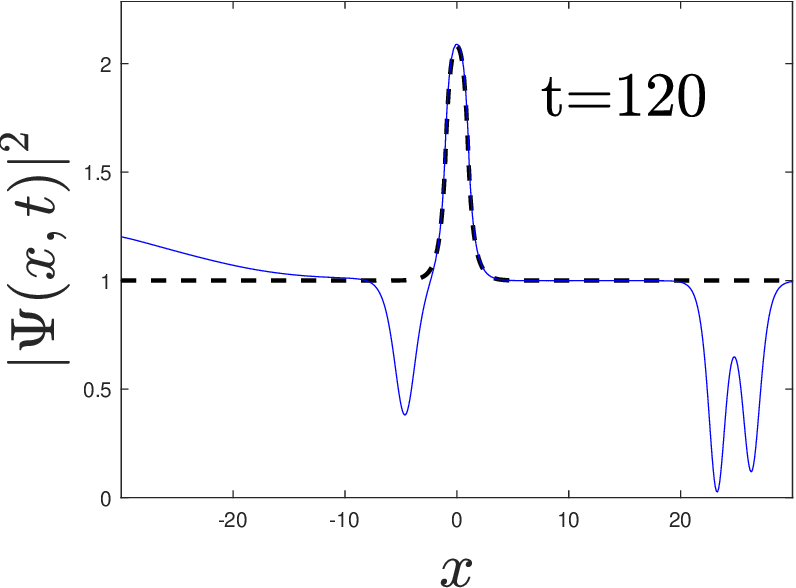} \\[0.5cm]
    \includegraphics[width=0.5\columnwidth]{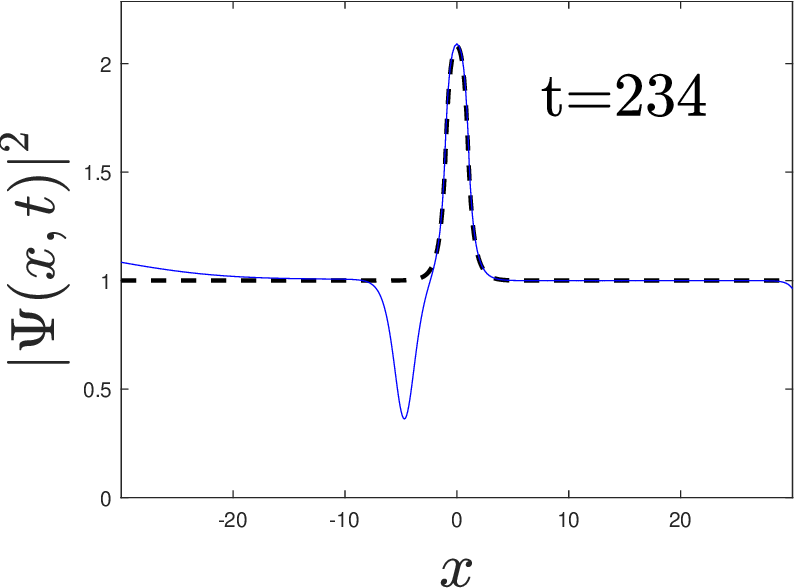} & \includegraphics[width=0.5\columnwidth]{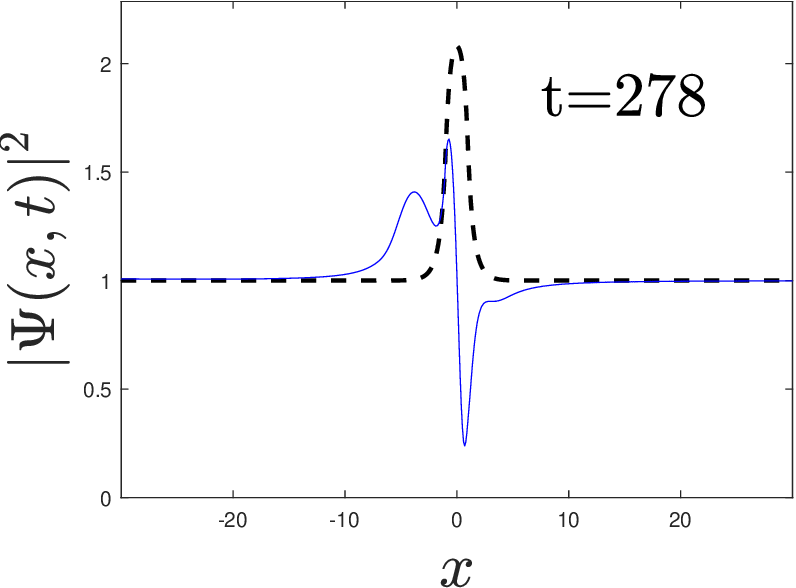} &
    \includegraphics[width=0.5\columnwidth]{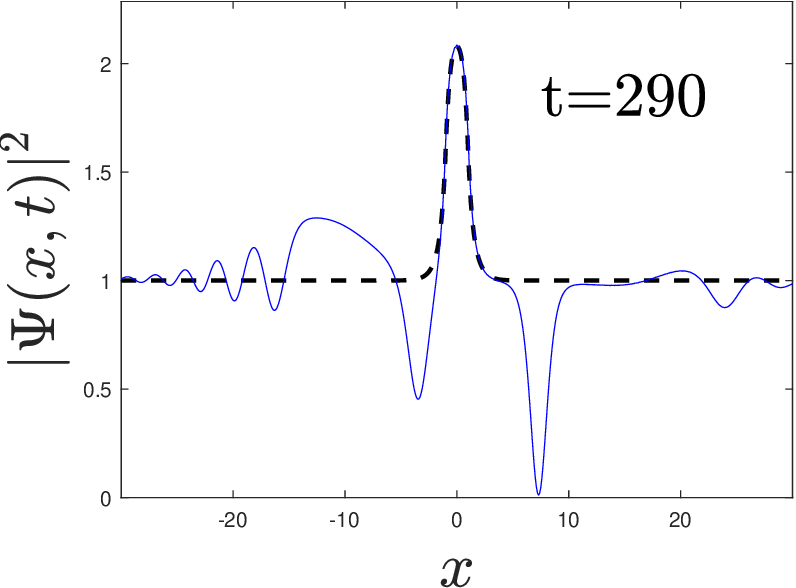} & \includegraphics[width=0.5\columnwidth]{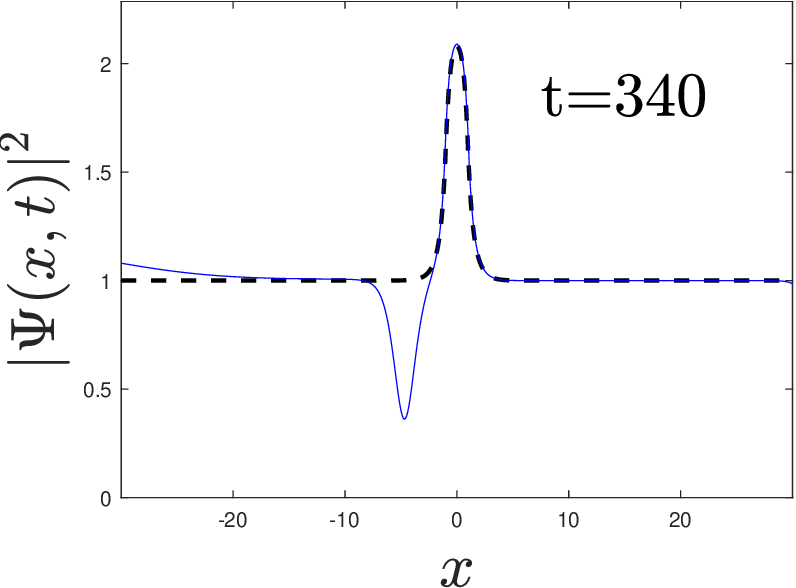} \\
\end{tabular}
\caption{Plot of the time evolution of the density profile $|\Psi(x,t)|^2$ (solid blue) for different times for the IHFC model. The ground state is depicted in dashed black line. Upper row: $V_0=1$, $X=2$ and $v=0.6$. Middle row: same as upper but for a configuration with $v=0.62$. Lower row: same as middle row but for later times, once the system is in the CES regime.}
\label{fig:CESTransition}
\end{figure*}

\section{Long-time phase diagram of the IHFC model}\label{sec:PhaseDiagram}

\begin{figure*}[!htb]
\begin{tabular}{lr}
    \includegraphics[width=\columnwidth]{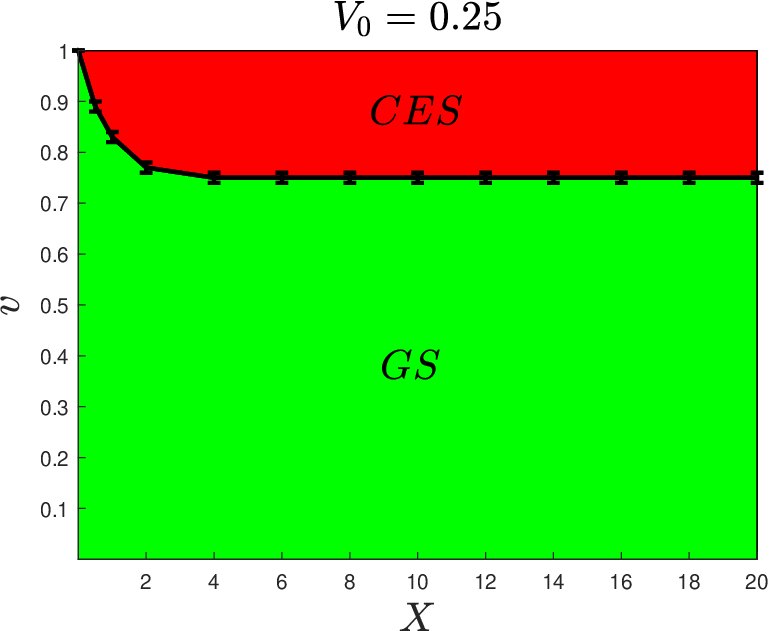} & ~~~~~
    \includegraphics[width=\columnwidth]{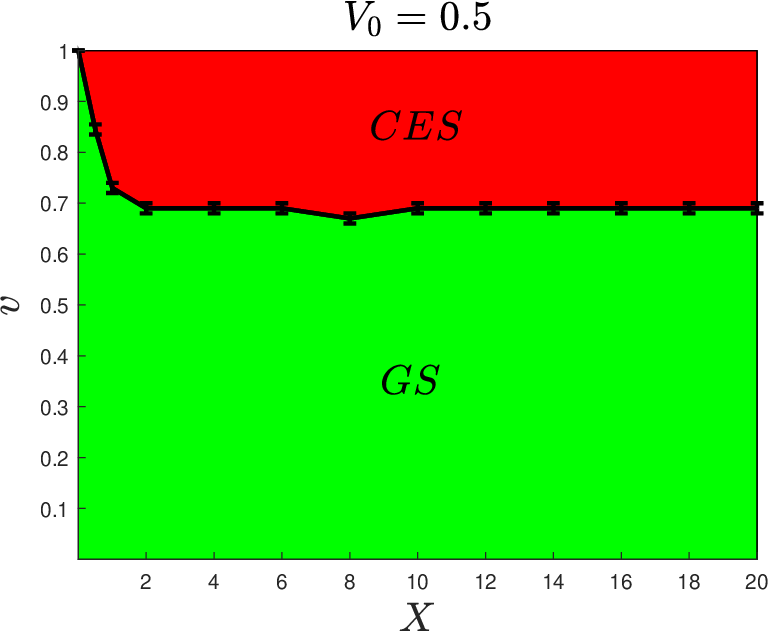}\\
    \includegraphics[width=\columnwidth]{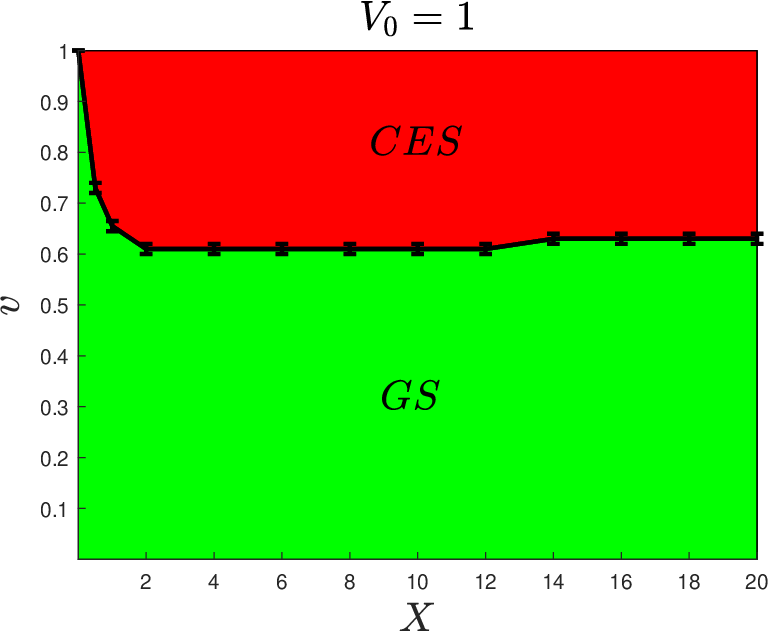} & \includegraphics[width=\columnwidth]{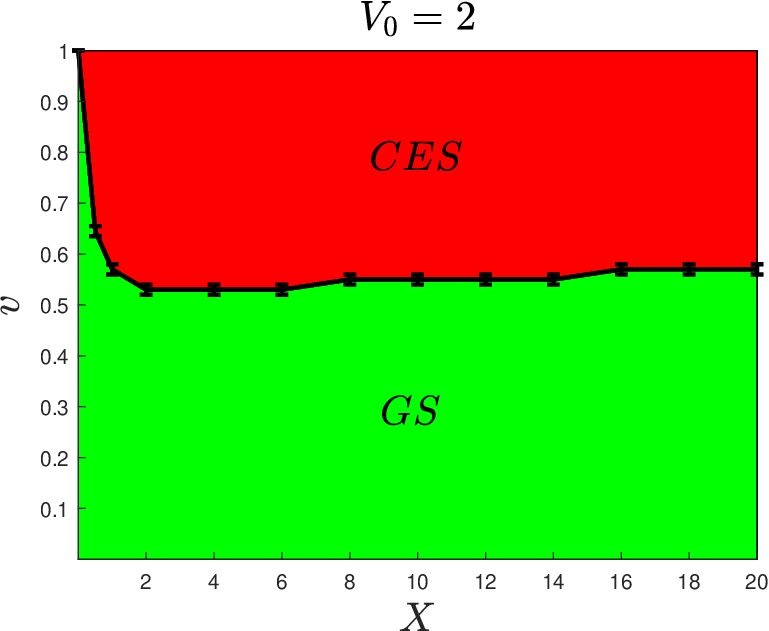}
\end{tabular}
\caption{Long-time phase diagram for the IHFC model. $GS$ denotes the region of phase space in which the system approaches the nonlinear ground state of the system and $CES$ that in which it enters a regime of continuous emission of solitons. The computation of the error bars is explained in the text. Upper left: phase diagram in the $(v,X)$ parameter space for $V_0=0.25$. Upper right: $V_0=0.5$. Lower left: $V_0=1$. Lower right: $V_0=2$.}
\label{fig:PhaseDiagram}
\end{figure*}


We study here a complementary model with no horizons, represented in Fig. \ref{fig:BHWFConfigurations}d, in which an attractive square well is suddenly introduced at $t=0$ in a subsonic homogeneous flowing condensate, characterized by a wave function $\Psi(x,0)=e^{ivx},~v<1$.  In this way, a time-dependent dynamics is induced since, in the new potential profile, such initial condition is not a stationary GP solution. We will refer to this setup as the IHFC model.

The idea is that the resulting Hamiltonian governing the time evolution of the IHFC model is exactly the same as for the generic BHL solutions, determined by the set of parameters $(v,X,V_0)$. The only difference is the initial condition at $t=0$, which now is a non-stationary GP solution that will follow a deterministic evolution. This contrasts with the case of a generic BHL solution, where the dynamics is stochastic, driven by the evolution of the initial random noise on top of the stationary BHL solution. Due to the strong insensitivity to the details of the transient reported in the previous section, one can expect the long-time phase diagram for the IHFC model to be identical to that for BHL solutions. Indeed, based on previous studies with localized defects \cite{Hakim1997}, scenarios of emission of trains of solitons are already anticipated in this kind of setup.

Several advantages motivate the introduction of this model. First, the values of $(v,X,V_0)$ can be chosen arbitrarily, without restriction. Second, since the ground state is smoothly connected to the initial homogeneous solution, the time required to explore the associated long-time dynamics is considerably shorter than for initial BHL solutions, which lowers the computational cost. Third, the evolution here is purely deterministic and there is no need to double check any sensitivity to the initial condition. Finally, we note that this model can be easily studied experimentally, and actually a similar setup was implemented in Ref. \cite{Engels2007}, in which a localized attractive potential was swept along a trapped condensate, observing emission of solitons.

As expected, in the numerical simulations it is observed that for long times the system only displays the same two possible behaviors. The transition between GS and CES for the IHFC model is shown in Fig. \ref{fig:CESTransition}, where in the first row we represent a case in which the system reaches GS at long times, and in the second and third rows we represent the same system but with a slightly higher flow velocity that puts the system in the CES regime. Interestingly, the mechanism giving rise to the CES state in the IHFC model is the same as that described in Fig. \ref{fig:BHLGenericTime}. We also remark that the times required to reach the long-time regime are considerably shorter than for initial BHL solutions; see Figs. \ref{fig:BHLFineTuneTime}, \ref{fig:BHLGenericTime}.

In Fig. \ref{fig:PhaseDiagram} we depict the numerical results for the long-time phase diagram of the IHFC model. The critical velocities $v_c\left(V_0,X\right)$ marking the phase boundaries are computed as follows: for fixed $X$ and $V_0$, the value of $v$ is changed. The critical value $v_c$ is computed as the average between a point in parameter space representing a simulation showing CES, and one evolving towards GS, also originating in this way the associated error bar, chosen of the order of the numerical precision. The critical value $v_c(V_0,0)=1$ is extracted from the Landau criterion of superfluidity \cite{Pitaevskii2003}. Indeed, this phase diagram may be viewed as a generalized nonlinear version of the Landau criterion for this potential profile.

In that phase diagram we also observe that, for increasing values of the well width $X$, the critical velocity saturates, $v_c\left(V_0,X\right)\simeq v_c\left(V_0\right)$. In the left panel of Fig. \ref{fig:CriticalAmplitude} we further check this feature by representing the values of the critical velocity $v_c\left(V_0,X\right)$ as a function of $V_0$ for $X=2$ (solid blue), $X=4$ (dashed red), and $X=10$ (dashed-doted green). We see that the function $v_c$ depends on $X$ very mildly since the three curves almost coincide within numerical precision.

\section{Robustness and universality of the long-time phase diagram}\label{sec:Robustness}

\begin{figure*}[!htb]
\begin{tabular}{lr}
    \includegraphics[width=\columnwidth]{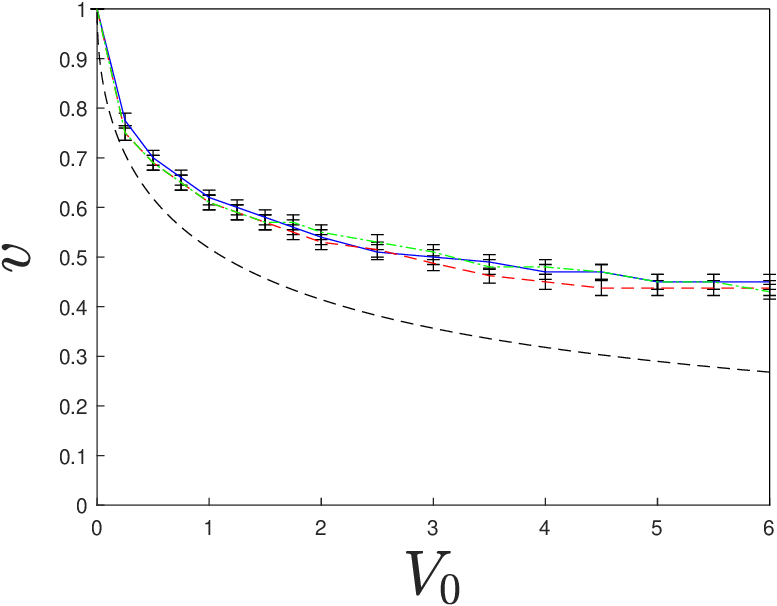} & ~~~~~
    \includegraphics[width=\columnwidth]{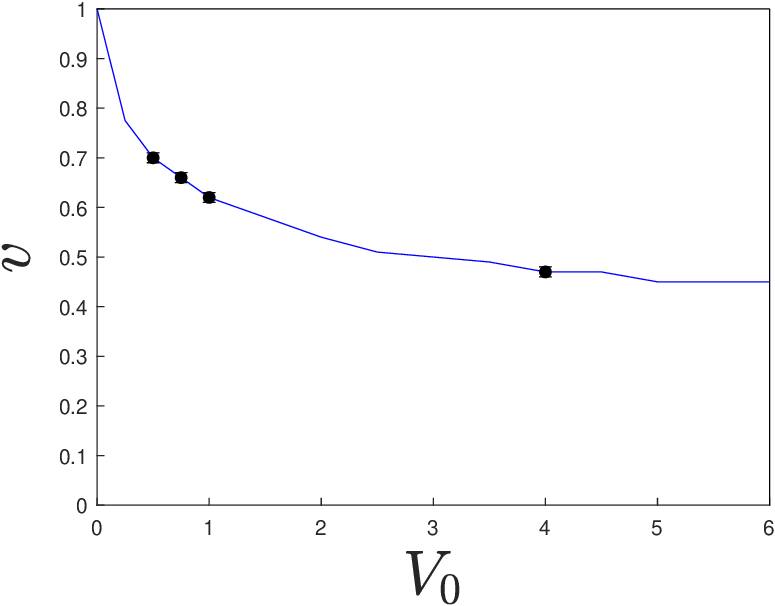} \\
\end{tabular}
\caption{Critical velocity $v_c\left(V_0,X\right)$, as a function of $V_0$, of the IHFC model. Error bars are computed as in Fig. \ref{fig:PhaseDiagram}. Left: $v_c\left(V_0,X\right)$ for $X=2$ (solid blue), $X=4$ (dashed red) and $X=10$ (dashed-doted green). Black dashed line marks the fine-tuned values $V_0(v)$ from Eq. (\ref{eq:waterfall}). Right: solid blue curve is the same as in left, black dots represent the critical velocities for initial generic BHL solutions with the same parameters $(v,V_0,X)$. The (small) error bars in the plot correspond to the numerical determination of the black dots.}
\label{fig:CriticalAmplitude}
\end{figure*}

We now connect the results for the long-time behavior of the IHFC model of Sec. \ref{sec:PhaseDiagram} with those of the BHL solutions of Sec. \ref{sec:TimeEvolution}. As a first step, we represent in the left panel of Fig. \ref{fig:CriticalAmplitude} the curve $V_0(v)$ characterizing the fine-tuned BHL configuration as a dashed black line, observing that it is well below the critical velocities predicted by the IHFC model. This observation, along with the found independence of the long-time behavior from the transient, provides an explanation of the reported lack of CES for initial fine-tuned BHL solutions.

Moreover, when available, we compare the critical velocities for initial generic BHL solutions with those for the IHFC model. An example of this comparison is provided in the right panel of Fig. \ref{fig:CriticalAmplitude}, where we plot the critical curve for $X=2$ for the IHFC model along with the critical velocities obtained using generic BHL solutions with the same background parameters as initial condition, finding a perfect agreement within the numerical uncertainties. The scarcity of points is due to the fact that, as explained, there are regions in the parameter space $(v,X,V_0)$ where there are no BHL solutions. Specifically, the comparison can only be carried out if a BHL solution is present close to the critical velocity $v_c(V_0,X)$. Since $v_c(V_0,X)\simeq v_c(V_0)$, the availability of a BHL solution at $(v_c(V_0),V_0,X)$ for fixed $X$ displays a structure in terms of allowed bands and forbidden gaps as a function of $V_0$. The black dots in the right panel of Fig. \ref{fig:CriticalAmplitude} then correspond to parameters $(v_c(V_0),V_0,X)$ within allowed bands, while the rest of the points of the blue curve are placed within gaps where no BHL solutions are available. For completeness, we have extended the IHFC-BHL comparison for other compatible values of $V_0$ and $X$ such that $(v_c(V_0),V_0,X)$ is within an allowed band, and in all cases a similar agreement for the critical velocities is observed (not shown).

Finally, we have also studied the role of the time scale of the switch-on of the potential in the IHFC model, allowing for a more gentle introduction of the square well instead of a sudden quench. We have found that the results for the long time phase-diagram are strongly independent of the transient time scale. The only exception is that, if the switch-on is sufficiently slow, the system always evolves towards the ground state, something that can be anticipated from the standard considerations on adiabaticity.

These results imply a very strong form of robustness for the long-time regime since two completely different types of evolution, one stochastic, driven by random noise added to unstable stationary black-hole laser solutions, and another one purely deterministic, arising from the time-evolution of a homogeneous initial condition, converge to the same state for long-times. The particular realization depends solely on the background parameters $(v,X,V_0)$ characterizing the global flow. This form of robustness goes even beyond that discussed in Ref. \cite{deNova2016}, which only involved linear random noise on top of the initial BHL configuration and a sudden quench in the coupling constant and external potential.

The observed robustness suggests that the CES regime is an intrinsic state of a flowing condensate and not a peculiar feature associated to some particular transient or phenomenon. Moreover, the IHFC configuration, which can be easily reproduced in the laboratory, could be used to study the long-time phase diagram of a BHL without the need to actually start from a true BHL configuration, whose evolution in current experiments is dominated by the background Bogoliubov-Cherenkov pattern originated at the inner horizon \cite{Kolobov2021}. Nonetheless, due to the strong insensitivity to the transient details, even in the actual experiment we can expect the resulting long-time nonlinear dynamics to be dominated by the same kind of nonlinear solutions and phase diagram here described.

In addition, the fact that the long-time behavior of the realistic BHL model here analyzed is the same as that of the idealized flat-profile BHL model bespeaks universality for the final fate of a black-hole laser, with only two possible choices: either the system reaches the ground state by evaporating away the horizons or it enters the CES regime, self-oscillating periodically while radiating solitons.

\section{Conclusions and outlook}\label{sec:conclu}

In this work, we have studied the time evolution of a realistic model of a black-hole laser. We have confirmed and extended results from previous works dealing with the idealized flat-profile configuration. We have computed the spectrum of linear instabilities and found a behavior similar to that described in the literature, except for the appearance here of a novel mode that makes the system unstable for any nonzero cavity length. In the nonlinear regime, there is also a phase diagram for the long-time behavior in which the system either goes to the ground state or to the CES state, suggesting a form of universality for the long-time behavior of a black-hole laser. Actually, the CES state provides the closest counterpart of an actual optical laser device, as discussed in Ref. \cite{deNova2016}.

In order to reach a deeper understanding of the long-time dynamics, we have developed a complementary deterministic model, based on the sudden introduction of an attractive square well within a homogeneous subsonic flowing condensate, with no horizons present. We have studied the associated long-time phase diagram and observed a perfect agreement with that of the corresponding black-hole laser configurations for the same background flow. This implies a stronger form of robustness with respect to the transient that goes beyond the BHL scenarios described in the literature.

From the gravitational analog perspective, the unambiguous observation of the black-hole laser effect still represents an ongoing challenge in the field. The present work provides a description of the genuine properties of the black-hole laser dynamics as we are able to isolate it in our model from the Bogoliubov-Cherenkov background. Thus, the results here reported can serve as a basis for future theoretical studies trying to identify a distinctive signature of black-hole lasing in an actual experimental setup. Furthermore, since we have observed that the long-time regime is quite independent of the transient details, we also expect that the results of this work can be useful for the study of the nonlinear behavior of the stimulated Bogoliubov-Cherenkov wave.



Now that the spontaneous Hawking effect has been finally observed \cite{Steinhauer2016,deNova2019,Kolobov2021}, an important next step is the observation of the backreaction of the emitted Hawking radiation, which in a real black hole causes its evaporation, a process studied within a semiclassical framework. In a condensate, several techniques are available to get insight into the backreaction of quantum fluctuations, in particular the well-known Truncated Wigner method \cite{Sinatra2002}, which in turn can be also regarded as a semiclassical approximation to the full quantum field dynamics \cite{Polkovnikov2003}. At $T=0$, the dynamics of a stationary black-hole laser is purely stochastic, driven by dynamical instabilities triggered by vacuum fluctuations. Actually, such a black-hole laser configuration behaves as a quantum unstable harmonic oscillator \cite{Finazzi2010}. Preliminary attempts to study backreaction in a black-hole laser with the help of the Truncated Wigner method can be found in Refs. \cite{Jain2007,Burkle2018}. Other scenarios where vacuum fluctuations act as seed of dynamical instabilities have been also successfully investigated with the help of the Truncated Wigner method \cite{Ferris2008,Shrestha2009}. Some general discussion on the power and the limitations of this method can be found in Refs. \cite{Sinatra2002,VanRegemortel2017} and, more specifically for the backreaction phenomena, in Ref. \cite{Butera2020}.

The amplification of the quantum fluctuations produced in a black-hole laser provides an ideal scenario where to start investigating backreaction effects in analog gravity: even though the full nonlinear dynamics of a condensate, described by the Gross-Pitaevskii equation, is very different to that of a real black hole, described by Einstein equations, and the lasing amplification process strongly modifies the statistics of fluctuations, making them more classical, still we can expect that the analog system can inspire some ideas for the final fate of a real astrophysical system. For instance, one may wonder what is the black-hole equivalent of the long-time phase diagram.

From a more general perspective, the current work provides a new scenario of emission of solitons, of interest not just within the condensed matter community \cite{Hakim1997,Burger1999,Denschlag97} but also for other fields such as nonlinear optics \cite{Mollenauer1980,Cundiff1999} or quantum fluids of light \cite{Amo2011,Sich2012,Opala2018}, where similar nonlinear equations of motion are satisfied. In addition, the CES state represents an attractive system for the fields of quantum transport and atomtronics \cite{Seaman2007,Panday2019}.

Several possible lines of research follow the results of this work. For instance, the strong robustness of the CES regime here described suggests that it is an intrinsic state of the background flow. Therefore, a detailed study of its properties is in order. Future work should also address the effect of quantum fluctuations on the evolution of a black-hole laser, devoting special attention to the backreaction and the stability of the CES state \cite{deNova2020a}.


\acknowledgments

We would like to devote this work to the memory of Renaud Parentani, with whom we enjoyed valuable discussions, especially during the visit of one of us (JRMdN) to Universit\'e Paris-Saclay. We also thank Florent Michel for useful remarks during the mentioned visit. This work has been supported by Grant FIS2017-84368-P from Spain's MINECO. IC acknowledges support from the European Union's Horizon 2020 Research and Innovation Program under grant agreement No 820392 (PhoQuS) and from the Provincia Autonoma di Trento.

\bibliographystyle{apsrev4-1}
\bibliography{Hawking}

\end{document}